\documentclass[titlepage,12pt]{article}
\usepackage{amssymb,amsmath,color,graphics,amscd,epsf,indentfirst,amsfonts}
\usepackage{epsfig}
\usepackage[dvipdfm,hypertex]{hyperref}
\usepackage{nicefrac}
\usepackage{scalefnt}
\usepackage[titles]{tocloft}
\usepackage{sectsty}
\allsectionsfont{\bf \scalefont{.7} \selectfont}
\subsectionfont{\bf \scalefont{.85} \it \selectfont}
\subsubsectionfont{\bf \scalefont{1} \it \selectfont}
\usepackage[T1]{fontenc}
\usepackage{lmodern}
\usepackage{bibspacing}

\def\blfootnote{\xdef\@thefnmark{}\@footnotetext}

\long\def\symbolfootnote[#1]#2{\begingroup%
\def\thefootnote{\fnsymbol{footnote}}\footnote[#1]{#2}\endgroup}

\makeatletter
\renewcommand{\@dotsep}{4.5}
\makeatother

\def\be{\begin{equation}}
\def\ee{\end{equation}}

\makeatletter
\def\@seccntformat#1{\csname the#1\endcsname.\quad}
\makeatother

\def\clock{{\count0=\time
           \divide\count0 60
           \ifnum\count0<10 0\fi\the\count0
           \multiply\count0 -60 \advance\count0 \time
           :\ifnum\count0<10 0\fi \the\count0
         }}
\newcommand{\timestamp}{{\small\vbox{\hbox{\tt\jobname.tex}
\hbox{\the\day/\the\month/\the\year, \clock}}}}

\def\sp{\;\;\;,\;\;\;}

\def\NN{{\cal N}}
\def\OO{{\cal O}}

\def\SS{{\cal S}}

\def\d{{\partial}}

\newcommand{\beq}{\begin{equation}}
\newcommand{\eeq}{\end{equation}}
\newcommand{\ba}{\begin{array}}
\newcommand{\ea}{\end{array}}
\newcommand{\bea}{\begin{eqnarray}}
\newcommand{\eea}{\end{eqnarray}}

\newcommand{\tr}{\mathop{{\rm Tr}}}

\def\hre#1#2{\href{http://arxiv.org/abs/#1/#2}{[ArXiv:#1/#2]}}

\def\hri#1#2{\href{http://arxiv.org/abs/#1}{[ArXiv:#1]#2}}

\numberwithin{equation}{section}

\begin{document}

\begin{titlepage}
\begin{flushright}
CPHT-RR 041.0608\\
\end{flushright}
\vskip 2.8cm
\begin{center}
\font\titlerm=cmr10 scaled\magstep4
    \font\titlei=cmmi10 scaled\magstep4
    \font\titleis=cmmi7 scaled\magstep4
    \centerline{\titlerm
      Interacting String Multi-verses and
      \vspace{0.4cm}}
    \centerline{\titlerm
      Holographic Instabilities of Massive Gravity}
\vskip 1.5cm
{\it Elias Kiritsis$^{\spadesuit,\diamondsuit}$ and Vasilis Niarchos$^{\spadesuit}$}\\
\vskip 0.7cm
\medskip
{$^{\spadesuit}$Centre de Physique Th\'eorique, \'Ecole Polytechnique,
91128 Palaiseau, France}\\
{Unit\'e mixte de Recherche 7644, CNRS}\\
\smallskip
{$^\diamondsuit$Department of Physics, University of Crete, 71003 Heraklion, Greece}

\end{center}
\vskip .4in
\centerline{\bf Abstract}

\baselineskip 20pt
%

\vskip .5cm \noindent
Products of large-$N$ conformal field theories coupled by multi-trace interactions
in diverse dimensions are used to define quantum multi-gravity (multi-string theory)
on a union of (asymptotically) AdS spaces. One-loop effects generate a small
${\cal O}(1/N)$ mass for some of the gravitons. The boundary gauge theory and
the AdS/CFT correspondence are used as guiding principles to study and draw
conclusions on some of the well known problems of massive gravity -- classical
instabilities and strong coupling effects. We find examples of stable multi-graviton
theories where the usual strong coupling effects of the scalar mode of the graviton
are suppressed. Our examples require a fine tuning of the boundary
conditions in AdS. Without it, the spacetime background backreacts in order
to erase the effects of the graviton mass.

\vfill
\noindent
August 2008
\end{titlepage}\vfill\eject

\setcounter{equation}{0}

\pagestyle{empty}
\small
\tableofcontents
\normalsize
\pagestyle{plain}
\setcounter{page}{1}

\section{Introduction}
\label{sec:intro}

\vspace{-.2cm}
\subsection{Massive gravity -- motivation and problems}
\label{sec:massive}

Gravity has been the first fundamental interaction to be known to humans and, as it
turns out, the hardest to really understand. Most of the major problems of fundamental
physics have an intimate connection with the problems of gravity. One of the most
important ones, the cosmological constant problem, is particularly acute today, especially
since we are becoming more and more certain that the observable universe accelerates
at a rate compatible with a vacuum energy density approximately equal to
$(10^{-27}M_P)^4$.

There are not many ideas that explain this fact, and those that have been proposed have
serious fine tuning problems. Such ideas attempt to modify gravitational dynamics
in the infrared (IR), in order to effectively screen the gravitational interaction.

One of the simplest ways to screen the gravitational interaction is to give the graviton a mass,
and this possibility was explored as early as the Fierz-Pauli paper \cite{fp}. It has been
further realized, following several important works of S.\ Deser and collaborators, that
gravitational theories with a mass term are behaving quite unlike most other theories.
In particular, they are generically unstable at the non-linear level. Gravity being a theory
with local gauge invariance invites us to suspect that maybe some of these problems
are due to the fact that the mass is introduced in a way that breaks gauge invariance
badly.

A graviton with a small mass has important cosmological implications.
Using the cosmological equations for massive gravity \cite{gb} one finds at late
times, while the universe is expanding, that there is an effective cosmological
acceleration, which is equivalent to a constant {\em positive} vacuum energy
$\Lambda^4\sim m_g^2M_P^2$ \cite{review}. If the mass of the graviton is of the order
of the inverse Hubble scale today, $m_g\sim H_0^{-1}$, then $\Lambda\sim 10^{-3}$
eV in agreement with todays' observations. This cannot be the whole story however.
It turns out that higher terms in the graviton potential give more and more dominant
contributions to the late time evolution of the universe, \cite{review}, thus indicating
that more needs to be understood in order for massive gravitons to provide a credible
explanation of todays' cosmological acceleration.

More recent investigations starting with \cite{ags}, implemented a Stuckelberg formalism
and observed that the Fierz-Pauli theory exhibits a strong coupling problem at energies that
can be hierarchically smaller than the Planck scale. To be more precise, if the graviton
mass $m_g\ll M_P$, then the theory becomes strongly coupled at energies comparable
or larger than $\Lambda_{V}=(m_g^4M_P)^{1\over 5}$, a scale that is hierarchically
smaller than the Planck scale. The interactions that become strongest are those of the
scalar mode of the graviton. The strong coupling problem was implicit in an older
observation of Vainshtein \cite{vain}, who studied the analog of the Schwarzschild
solution (and its breakdown) in the Pauli-Fierz theory. When external fields (due to
masses) are strong, they can also trigger the coupling to be strong and the perturbative
expansion to break down. Vainshtein found that the linearized solution of a point mass
$M$ broke down at a distance
\be
R_{V}\sim \left({M\over M_P^2 m_g^4}\right)^{1\over 5}
\ee
from the source. By judiciously modifying the graviton potential \cite{ags} this bound can
be improved to
\be
R_{AGS}\sim \Lambda^{-1}\left({M\over M_P}\right)^{1\over 3}
\ee
where $\Lambda$ is the ultraviolet (UV) cutoff, or equivalently the energy scale at which
the interactions become strong, given in (\ref{aaa}).

The fact that gravity becomes strong would pose, a priori, no major problem
at the classical level since there are several cases where the full non-linearity of the theory
can be handled. For example, this is the case of Einstein gravity or its Lovelock
generalizations where the full non-linear action is known. However, it is the
quantum (loop) effects that cannot be neglected in this regime and pose a problem.
This is because, typically, the UV quantum structure of massive graviton theories
is essentially unknown. Therefore, the scale $\Lambda$, where the scalar graviton
mode interactions become strong, is a UV cutoff in the standard sense of quantum
field theory. Beyond this energy one must know the UV completion of the theory
in order to proceed. This peculiar IR behavior of massive gravity is reflecting an
essential feature: the failure of IR physics to decouple in a simple way from UV
physics. The theory has two characteristic length scales, $m_g$ and $M_P$,
however, there is a dynamically generated intermediate scale $\Lambda_V$
where the (quantum) theory breaks down, or the classical gravitational interaction
becomes strong. This appears to be an important loophole for many IR approaches
to quantum gravity which invoke a complete decoupling of UV and IR effects.

Since the Einstein term in gravity is a two-derivative term, any local IR modification
of gravity proceeds via a potential for the graviton. A graviton mass is the leading
non-trivial part of this potential, but higher order terms can be present.\footnote{It has
been suggested in \cite{kd1,kd2} that it is possible to have a graviton potential which
starts at the cubic level. This, however, does not seem possible due to the principles
of spontaneously broken diffeomorphism invariance.} It seems that the peculiarities
of the mass term and its non-decoupling properties are shared by other terms in the
potential as well. On the other hand, as shown in \cite{ags}, by fine-tuning the graviton
potential, the scale where gravity becomes strong can be improved to
 \be
 \Lambda_{\rm AGS}=(m_g^2M_P)^{1\over 3}\;\;.
\label{aaa} \ee

To visualize the size of this cutoff scale, in the case where we give the graviton a
mass of the order of the inverse size of the observable universe today ($\sim 10^{28}$ m)
$m_g\sim 10^{-34}$ eV, we obtain $\Lambda_{V}\sim 10^{-22}$ eV corresponding to a
distance scale of $10^{13}$ Km, or about 1 light year. Therefore gravity for distances
smaller that 1 light year would be strongly coupled. The same estimates would still give
$\Lambda_{\rm AGS}\sim 10^7$ m for the improved cutoff, comparable with the
earth-moon distance. Only a cutoff that would scale as $\Lambda_*\sim \sqrt{m_gM_P}$
would translate into a distance scale of a millimeter, which is marginally compatible with
todays' experimental results on the gravitational force showing that down to a fraction
of a millimeter Newton's law is valid. At shorter distances the theory is strongly coupled
and a UV completion is necessary. A massive graviton theory with $\Lambda_*$ as a
cutoff is not known currently. We will show in this paper that gravity (string)
theories in (asymptotically) AdS spaces coupled via boundary interactions do
realize (in four spacetime dimensions) massive graviton theories with $\Lambda_*$
as the cutoff and at the same time possess a complete UV description in terms of a
dual large-$N$ field theory.

Another characteristic feature of massive graviton theories is the Van Dam-Veltman-
Zakharov (vDVZ) discontinuity \cite{vvd}, $i.e.$ the fact that as we send
$m_g\to 0$ the scalar graviton mode fails to decouple. This is of course
correlated with the fact that around flat space the scalar graviton interactions
have a singular limit as $m_g\to 0$ and that the associated UV cutoff asymptotes
to zero. It should be noted, however, that this is a background dependent feature.
For example, around AdS backgrounds this discontinuity does not exist \cite{porrativvd}
and as we will see later, the cutoff estimates typically change.

To formulate a massive gravity theory, or to write down a general potential for the
graviton, a fiducial metric must be introduced. This metric is non-dynamical and is
typically taken to be flat, although constant curvature metrics are also considered.
A natural theoretical generalization of the theory involves making the fiducial metric
dynamical. In this case, we are dealing with a bi-gravity theory,\footnote{Bi-gravity was
first introduced in \cite{salam} in the context of the strong interactions. See
\cite{kd1,kd2,kd3} for a recent detailed classical study of such theories and their
solutions. } a theory with two propagating gravitons, two in general distinct
Planck scales and in the decoupling limit, two diffeomorphism invariances.
By adding a potential that couples the two gravitons together it is possible to break
one of the diffeomorphism invariances and give a mass to a linear combination
of the two gravitons. Taking one of the Planck masses to infinity freezes the dynamics
of one of the metrics and we return to the previous type of massive theories in a
fixed fiducial metric. Bi-gravity theories naturally generalize to multi-gravity theories
involving more than two metrics \cite{kd1,grosj}.

It has been shown in \cite{deser}, and more recently in \cite{insta}, that massive
graviton theories are generically unstable. This is a non-linear effect that is generic,
and has its source in the involved non-linear structure of the kinetic terms of the
graviton action. The instability arises in the perturbative spectrum as a ghost
mode (occasionally this mode may also be a tachyon).
Although some times in a {\em fixed background} the instabilities
can be fine-tuned away, small departures from the original background are
enough to reinstate them. Even in cases where one considers non-Lorentz
invariant potentials as in \cite{ghost}, and several other cases reviewed in
\cite{rreview}, the theory is stable only in a subclass of backgrounds. Unlike
the case of conventional symmetries, here symmetries cannot prohibit
non-symmetric fluctuations and therefore the instabilities are real. In some cases
it may be argued that it will take long for the instabilities to be felt in our universe
\cite{rreview}.

It is important to recall here an insight from the first of references \cite{insta} stating
that the strong coupling cutoff and the non-linear instability are intrinsically linked.
Both of these features have a common source in the high-derivative couplings of the scalar
graviton mode, which induce, on one hand, instabilities characteristic of high derivative
theories and, on the other hand, deteriorate the power counting of the associated
interactions.

The above behavior is not only characteristic of theories where a graviton has a mass
term in the Lagrangian, but also more exotic cases where the massive graviton
is a resonance. This is what happens, for instance, when gravity is induced on
branes, the most celebrated example being given by the DGP model \cite{dgp}.
Indeed, it has been observed in \cite{ktt} that this theory exhibits a similar
non-decoupling behavior as the standard Pauli-Fierz theory. It was subsequently
shown that the theory becomes strongly coupled at hierarchically low(er) scales,
\cite{lpr,rub}. There are also indications that the generic non-linear instabilities of
massive gravity also occur in brane-induced gravity as reviewed in \cite{Charmousis:2006pn,
Gregory:2007xy,greg}.

Another point to be stressed is the fact that the problems of massive gravity originate
in the self-interactions of the massive gravitons. Once other interactions are included
the conclusions mentioned so far may change. This becomes abundantly clear by
studying the case of KK compactifications of gravity \cite{sch}, which from the lower
dimensional point of view can also be thought as a theory with an infinite number of
interacting massive gravitons. What happens here is that the interactions of massive
as well as massless modes are tuned in a way that removes the strongly coupled interactions
and pushes the regime of validity of perturbation theory at scales that are much higher
than the naive cutoffs.

In view of the potential cosmological significance of theories with gravity modifications
in the IR as well as the fundamental theoretical limitations due to the existence of
rather low UV cutoffs, it is sensible to seek out theories of modified gravity which
are UV complete. In principle, one could use such examples to study both the
mechanisms of UV-IR mixing and how the theory behaves in the strong coupling
limit. Most importantly, one would like to know eventually whether such theories
have any chance of describing the experimental data.

\subsection{Massive gravity in a holographic context -- an overview of the paper}
\label{sec:overview}

What we will pursue in this paper is the case of a large class of massive
multi-graviton theories in diverse dimensions that are dual to large-$N$
quantum field theories (QFTs), which are product theories coupled together by a
multiple trace interaction. The prototype consists of two $d$-dimensional large-$N$
conformal field theories (CFTs), CFT$_1$ and CFT$_2$, each one having a dual
description in terms of a gravity (string/M) theory on AdS$_{d+1}$. It has been argued
\cite{kir,ack} that the dual description of the direct product theory CFT$_1\times$CFT$_2$
is a bi-gravity theory on a union of AdS spaces. Since there is no interaction between
the two CFTs, the dual theories on each AdS do not communicate with each other.
Hence, this is a trivial bi-gravity theory with two massless, non-interacting gravitons.
The only way to couple the two CFTs without spoiling the gauge symmetries of each
theory is by multi-trace interactions, $e.g.$ a double-trace interaction of the form
$g \OO_1\OO_2$, where $\OO_{1,2}$ are single-trace operators in CFT$_{1,2}$
respectively. A deformation of the latter form violates the conservation of the
energy-momentum tensor of each CFT, retaining only the conservation of
the total stress tensor. Since the energy momentum conservation in the boundary
is intimately related to diffeomorphism invariance in the bulk, we deduce that
the bulk theory must have a reduced group of diffeomorphism symmetries, hence
some of the gravitons must be massive.

Computationally, the mass of the gravitons comes about in the following way.
The effect of multiple trace deformations in the dual gravitational theory is well known
at tree-level \cite{witten,berkooz,muck,minces,petkou}. In the regime where we can trust
supergravity, the main effect is a change of the standard Dirichlet or Neumann
boundary conditions to mixed boundary conditions for the scalar fields that are
dual to the operators $\OO_i$ participating in the deformation. The precise form
of the mixed boundary conditions in our multi-AdS setup can be determined in
a straightforward manner by a simple generalization of the arguments of
\cite{witten,berkooz,muck,minces,petkou}.

\begin{figure}[t!]
\centerline{\includegraphics[width=7cm,height=7cm]{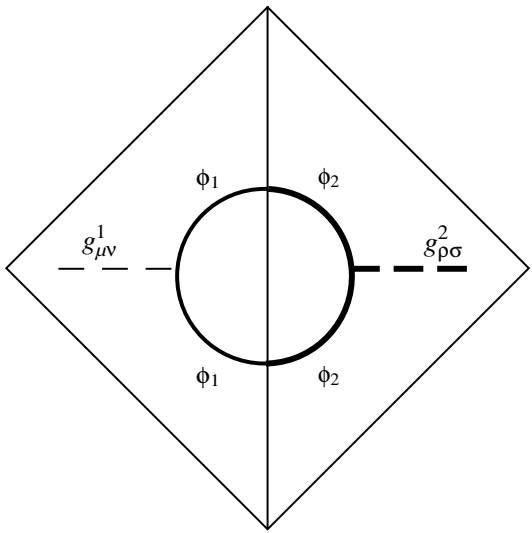}}
\caption{\small \it A diagrammatic representation of the one-loop amplitude that gives
a mass to a linear combination of two gravitons, each living on a different AdS space.
The mixed boundary conditions for the dual scalars $\phi_1,\phi_2$ give a
non-vanishing 1-2 graviton propagator.}
\label{fig:loop}
\end{figure}

References \cite{kir,ack} (see \cite{porrati} for related earlier work)
observed that the mixed boundary conditions facilitate a scalar loop correction
to the graviton propagator in the bulk of the form depicted in fig.\ \ref{fig:loop}.
This correction includes a non-zero mass term  for a  linear combination of the gravitons.
For a double-trace deformation of the form $g \OO_1\OO_2$ one determines
a mass of the order $g/N$ in accordance with the non-conservation of energy momentum
tensors on the boundary QFT. A discussion of the graviton mass matrix in a more general
context of a network of interacting CFTs generalizing the result of \cite{ack}
can be found in appendix \ref{app:genmass}.

In this paper we will revisit this holographic context. Our goal is to use it as a
laboratory to learn something new about the tantalizing questions of massive
gravity that were outlined in the previous subsection. The advantage we hope
to utilize is the extra power provided by the AdS/CFT correspondence.

In preparation for this task we begin in section \ref{sec:estimates} with a general
discussion of UV cutoffs in massive gravity in AdS spaces. Using various
scaling properties in a general set of large-$N$ AdS/CFT duals, and the
arguments of \cite{ags} for the strong coupling scales of massive gravity we arrive
at a universal minimal UV cutoff set by the AdS scale. In the process, we can also
find situations that realize the phenomenologically preferred cutoff $\Lambda_* \sim
\sqrt{m_gM_P}$, which is hard to obtain in other contexts. The  question we address
is whether this cutoff is a real signal of strong coupling effects or whether it is just a
mirage that does not reflect the real dynamics. To answer this question we will turn
to the QFT dual.

The main part of the QFT analysis is contained in section \ref{sec:higherD}.
We concentrate on a pair of $d$-dimensional CFTs deformed by a general
multi-trace operator that involves two single-trace operators, one from CFT$_1$
and another from CFT$_2$. By analyzing the renormalization group (RG)
equations in conformal perturbation theory, we find an interesting pattern
of perturbative fixed points with non-trivial interaction between the original
theories, provided that the scaling dimensions of the single-trace operators
participating in the deformation satisfy a certain compatibility condition.
Furthermore, we find that the fixed points are repellors of the RG equations.
Deforming away from them drives the theory towards a fixed point with
no interaction between the original CFTs or towards strong coupling
where our analysis in conformal perturbation theory breaks down.

With this understanding on the field theory side, we return in section
\ref{sec:lessons} to the massive multi-gravity theory in the bulk. We argue
that the low-energy effective action is a bi-gravity theory in the general
spirit of Damour and Kogan \cite{kd1} with some extra non-local features
whose origin can be traced to an underlying non-local string theory (NLST)
\cite{nonlocal}. Unfortunately, we do not have an explicit understanding of the
NLST rules in this context.\footnote{There is, however, a toy model
two-dimensional version of our theories, where the NLST rules can be
analyzed to all orders in perturbation theory \cite{kirniar1}.} Nevertheless,
there are some general properties of these theories that we can deduce from
the structure of the RG equations on the boundary QFT.

We conclude section \ref{sec:lessons} with an attempt to determine if our
massive graviton theories exhibit the usual problems of massive gravity, $i.e.$
instabilities and strong coupling problems. We argue against both of these
features and propose that our previous estimate of a strong coupling
around the AdS scale is indeed a fake -- extra interactions should be
taken into account which should lead to a softening of the ampilitudes.

Finally, section \ref{discussion} contains a summary of our main results
and some additional remarks on other interesting aspects of this work.
Several technical details have been relegated to four appendices at the
end of the paper.

\section{Large-$N$ theories and massive gravitons: estimating the cutoffs\label{estimate}}
\label{sec:estimates}

There are several large-$N$ conformal field theories that have been linked via an
AdS/CFT correspondence to gravitational (string/M) bulk theories. Here is a list of
such theories with a summary of their most important scales:
\begin{itemize}

\item[(1)] {\it Standard adjoint U(N) gauge theories.}
The prototype is four dimensional $\NN=4$ super-Yang-Mills theory. The bulk dual
is string theory on AdS$_5\times$S$^5$. The following scale relations are valid in this
case
\be
AdS_5: ~~~\ell_{AdS}\sim \ell_s(g_sN)^{1\over 4}\sp (M_5\ell_{AdS})^3\sim N^2
~,
\label{m40}
\ee
where $M_5$ is the five-dimensional Planck scale, $\ell_{AdS}$ the AdS radius,
$\ell_s$ the string length scale and $g_s$ the string coupling constant.
In general, we will assume a leading order gravitational action parametrized in $d$
dimensions as
\be
S_{\rm grav}=(M_d)^{d-2}\int d^dx \sqrt{g} ~ R
\label{m44}
~.
\ee
We expand around a classical solution $g=g^{(0)}+h$ and define a canonically normalized
fluctuation
\be
h_c=(M_d)^{{d-2\over 2}}~h
\label{m45}
~.
\ee
Hence, the $h^{2+n}$ interaction term scales as ${\cal O}\left(M_d^{-{d-2\over 2}n}\right)$.
For standard large-$N$ $U(N)$ gauge theory this scaling will match the $1/N$ expansion of
stress tensor correlation functions if $(M_d)^{d-2\over 2}\sim N$. This is indeed true in
$d=5$ corresponding to four dimensional large-$N$ $\NN=4$ super-Yang-Mills theory. In other
dimensions the gauge theories may not be conformal but the large-$N$ estimates
presented here still hold. This implies that the associated Planck mass should scale as
\be
{\rm Adjoint~ gauge~ theory}~~:~~~~~M_d\sim N^{2\over d-2}
\label{m46}
~.
\ee

\item[(2)] {\it M2-type large-$N$ theories.}
The prototype in this class is the world-volume theory on $N$ M2-branes. Although
we know little about this theory,\footnote{See, however, \cite{m2} and followup work
for recent progress in understanding this theory.} we understand somewhat its bulk-boundary
correspondence. The bulk geometry AdS$_4\times $S$^7$ provides the holographic
description of a three dimensional CFT with $SO(8)$ global $R$ symmetry. The relevant
scales are
\be
AdS_4:~~~ \ell_{AdS}\sim \ell_p~N^{1\over 6}\sp (M_4\ell_{AdS})^2\sim N^{3\over 2}
\label{m41}
\ee
where $\ell_p$ is the eleven-dimensional Planck scale. In this case the expansion
parameter of the connected stress-tensor $n$-point functions is, according to (\ref{m41}),
$M_4\sim N^{3\over 4}$. Therefore, the connected $(n+2)$-point function of the
stress-tensor and other operators is of order ${\cal O}\left(N^{-{3\over 4}n}\right)$.

In dimensions other than four, a similar estimate for the correlation functions
implies that the $d$-dimensional Planck scale must scale as
\be
{\rm M2-type ~ theories}~~:~~~~~M_d\sim N^{3\over 2(d-2)}
~.
\ee

\item[(3)] {\it M5-type large-$N$ theories.}
The prototype in this case is the world-volume theory on $N$ M5-branes.
Again we know little about this theory directly, but we can infer some properties
from the corresponding bulk-boundary correspondence. The bulk geometry
AdS$_7\times $S$^4$ describes holographically a six dimensional CFT with
$SO(5)$ global $R$ symmetry. The relevant scales are
\be
AdS_7:~~~ \ell_{AdS}\sim \ell_p~N^{1\over 3}\sp (M_7\ell_{AdS})^5\sim N^{3}
\label{m42}
~.\ee
In this case the expansion parameter of the connected stress-tensor $n$-point
functions is $(M_7)^{5\over 2}\sim N^{3\over 2}$.
Therefore, the connected $(n+2)$-point function of the stress-tensor and other
operators is of order ${\cal O}\left(N^{-{3\over 2}n}\right)$.

In dimensions other than 7, a similar estimate for the correlation functions implies
that the $d$-dimensional Planck scale must scale as
\be
{\rm M5-type~theories}~~:~~~~~M_d\sim N^{3\over (d-2)}
~.
\label{m43}
\ee

\end{itemize}

\vspace{-.2cm}
\subsection{Estimates of the graviton masses\label{gm}}

The one-loop generated graviton mass for the dual of adjoint gauge theories
was calculated in \cite{kir,ack} and the result is\footnote{Such formulae are more
generally valid in asymptotically AdS backgrounds. A discussion of the graviton
mass matrix in setups with more complex multi-trace interactions can be found
in appendix \ref{app:genmass}.}
\be
m\ell\sim {h\over N}
\label{m52}\ee
where $\ell$ is the AdS length scale, and $h$ the dimensionless $\OO(1)$ coupling
of the multiple trace deformation that couples a pair of CFTs (not to be confused
with the metric perturbation $h$ in the bulk). For fixed $h$,  in the
large-$N$ limit, we always have $m\ell \ll 1$. Attempting to take $h\sim \OO(N)$
will upset the large-$N$ expansion of the gauge theory.

The result \eqref{m52} can be generalized easily to the case of M2- and M5-type
theories. The key ingredient in this calculation is the realization that the mass
term is related to the product of two connected three-point functions. More
specifically, let us consider a boundary CFT of the form CFT$_1 \times$CFT$_2$.
We perturb the action of this theory by a double-trace operator $\delta S\sim \int h
~\OO_1\OO_2$, where $\OO_{1,2}$ are gauge invariant operators in CFT$_{1,2}$
respectively. The mixing of the two gravitons in the bulk is generated from the
non-vanishing of the correlator
\be
\langle T_1 T_2\rangle \sim h^2 \langle T_1 \OO_1\OO_1\rangle \,
\langle T_2 \OO_2\OO_2\rangle
\label{m53}
\ee
where $T_{1,2}$ are the stress-energy tensors of CFTs 1 and 2. This mixing term is
proportional to $m^2$. In the usual adjoint CFTs, the right hand side of \eqref{m53}
is of order  ${\cal O}\left({h^2\over N^2}\right)$, leading to the estimate (\ref{m52}).
A formula similar to (\ref{m53}) holds also for the M2 or M5 CFTs and leads to the
estimates
\be
{\rm M2-CFT}~~:~~~~m\ell\sim {h\over N^{3\over 4}}
\label{m54}
~,
\ee
\be
{\rm M5-CFT}~~:~~~~m\ell\sim {h\over N^{3\over 2}}
~.
\label{m55}
\ee
We stress again the obvious fact that in all of the above cases $m\ell\ll 1$ in the
large-$N$ limit.

\vspace{-.2cm}
\subsection{The associated UV strong-coupling cutoffs}
\label{UVstrong}

The usual theories of massive gravity exhibit a set of strong coupling scales
where the interactions of the Stueckelberg modes of the massive graviton
become strong. Here we summarize the relevant scales for massive graviton
theories in AdS$_d$ using the formalism of \cite{ags}. Further technical details
can be found in appendix \ref{apa}.

We will denote the UV cutoff scales as $\Lambda_{p,q}$. $\Lambda_{p,q}$ is
by definition the energy scale where an interaction of the form
\beq
\label{Ipq}
I_{p,q} \sim m^2 (M_d)^{d-2} (\d A)^p \, (\d^2\phi)^q
\eeq
becomes strong. $A$ and $\phi$ are the vector and scalar modes of the massive
graviton. We are suppressing the tensor structure in these couplings, since it
will not be important for our purposes.

There are two interesting regimes of parameters we would like to focus on:
\begin{itemize}
\item[$(i)$] {\it The ``flat space regime''}: $m\ell\gg 1$.
In this case, the mass gap lies in the region where the AdS curvature is invisible.
This is the reason why we call this regime the flat-space regime. For energies
$E\gg 1/\ell$ the massive gravity theory lives effectively in flat spacetime.
\item[$(ii)$] {\it The ``AdS regime''}: $m\ell\ll 1$.
In this case, the mass gap is in the AdS region where the AdS curvature is visible.
For energies $m\ll E\ll {1\over \ell}$ the massive graviton theory resides in a space
with a visible curvature, whereas in the regime $E\gg {1\over \ell}$ we are effectively
around flat space.
\end{itemize}

In the flat space limit, $m\ell\gg 1$, the analogue of the Vainshtein cutoff is
\be
\Lambda_{0,3}\sim M_d\left({m\over M_d}\right)^{8\over d+6}
~.
\ee
By judiciously arranging the interactions the cutoff can be raised to
\be
\Lambda_{2,1}\sim M_d\left({m\over M_d}\right)^{4\over d+2}
\ee
which is always larger than $\Lambda_{0,3}$ for $m\ll M_d$.
Furthermore, it is interesting to observe that in $d=10$,
$\Lambda_{0,3}=\Lambda_{*}\sim \sqrt{mM_d}$, while in
$d=6$,  $\Lambda_{2,1}=\Lambda_{*}\sim \sqrt{mM_d}$.
Recall that $\Lambda_*$ is a phenomenologically appealing
scale, which is marginally compatible with todays' experimental
results on the gravitational force.

The case, however, which is most relevant to the large-$N$ theories
that we study in this paper, is the case with a graviton mass that is
much smaller than the AdS length, $i.e.$ $m\ell\ll 1$. As we saw in the
previous section \ref{gm}, the combination $m\ell$ is
always suppressed by a positive power of $N$.

To proceed further we need to relate the scales $m, M_d$ and $\ell$.
The precise relation could depend on the type of large-$N$ theory we consider
within the classification in the beginning of this section. In all cases, however,
we obtain a universal strong coupling cutoff of the order of the AdS scale
(see appendix \ref{apa})
\be
\Lambda_{min}\sim \Lambda_{0,2}\sim \Lambda_{p,\infty}\sim {1\over \ell}\sim
{1\over h^{2\over d}} \left(m~M_d^{d-2\over 2}\right)^{2\over d}
~.
\ee

Therefore we seem to conclude that the massive graviton theory, as an effective
low energy theory, would be unreliable at distances shorter than the AdS scale.
Also, note that for $d=4$ we obtain
\be
\Lambda_{min}\sim {1\over \ell}\sim {1\over \sqrt{h}}\sqrt{m~M_4}\sp m\ell\sim
\sqrt{h}\sqrt{m\over M_4}
\label{m62} \ee
and as a result the cutoff is hierarchically larger than
$\Lambda_*\sim  \sqrt{m~M_4}$ for $h\ll 1$.

The above discussion assumes that the only interactions are those of the
massive graviton itself. It is possible, however, that in the presence of other
interactions -- and, indeed, many more interactions are present in our setup --
the strong coupling effects are substantially smoothed and that
the above cutoffs are merely a ``mirage''. There are two well known examples
where this is actually the case.

The first example concerns massive Kaluza-Klein (KK) gravitons. Starting from
a gravity theory in $D$ dimensions with Planck scale $M_D$, we compactify
$D-d$ dimensions on a torus, which, for simplicity, we will take to have a uniform
radius $R$, much longer than the fundamental Planck length, $i.e.$ $M_DR\gg 1$.
Hence, the lighest KK graviton mode has a mass $1/R$. Estimating the strong
coupling cutoffs associated with this graviton KK mode and assuming no more
than 4 compact dimensions we find (see appendix \ref{app:KKgravitons})
\be
{1\over R}\ll  \Lambda_{0,3}\ll \Lambda_{2,1}\ll M_D\ll M_d
~.
\label{m777}
\ee
The cutoffs are below $M_D$. Since we know that the theory is well defined
with a $D$-dimensional massless graviton in that regime, we are forced to conclude
that the estimated strong coupling scales are ineffective, merely a mirage of
our naive estimate. Indeed, it was shown explicitly in \cite{sch} that the strong
coupling effects delicately cancel out in diagrams due to the exchange of both
massless fields and the second level KK graviton states.

Another instructive example concerns massive spin-two fields that are oscillations
of a string. Most string theories contain an infinity of such states, the lowest
massive ones having a mass of order the string scale $M_s$. In the case of
ten-dimensional string theories in the perturbative regime around flat space,
a naive estimate of the strong coupling cutoffs associated to a stringy massive
graviton mode gives (see appendix \ref{app:stringgravitons})
\be
m\sim M_s \ll \Lambda_{0,3}\ll \Lambda_{2,1}\ll M_{10}
~.
\label{m711}
\ee
We know, however, that the string perturbation theory is well defined at
any scale hierarchically below the Planck scale, and therefore we deduce that
the cutoffs  $\Lambda_{0,3}$ and  $\Lambda_{2,1}$ are again a mirage of our estimate.
There is no associated strong coupling problem, most probably via judicious cancelations
similar to those observed in the KK case \cite{sch}.

In the holographic cases of interest in this paper the picture that
seems to be emerging is the following. First, we are always in the
AdS regime of parameters, where the one-loop generated graviton
mass $m$ is much smaller than the AdS energy scale $1/\ell$. A naive
estimate of strong coupling scales, based on self-interactions of the
graviton, indicates that the low-energy effective theory will break down
universally around the AdS energy scale. This would seem to imply,
in particular, that we cannot access the flat space regime of energies
$E$ beyond the AdS scale $1/\ell$.

Our system, however, is a complicated theory of gravity with many fields
besides the metric and many interactions that the above analysis of strong
coupling scales did not account for. In fact, our theory is not even expected to
be a conventional theory of supergravity -- as a low-energy effective
description of a non-local string theory it is expected to have extra
non-local features, some of which will be discussed further in section
\ref{sec:lessons}. Hence, as suggested by the previous examples,
it is highly possible that the above estimated strong coupling scales
are merely a mirage and that the theory is well defined all the way up
to the string scale. Evidence pointing in this direction is also provided
by the dual gauge theory description that will be discussed in the next
two sections.

For this picture to be correct it is important that the extra physics, responsible
for the smoothing of the interactions, come from fields that have masses
at or below the estimated strong coupling cutoff. Indeed, the cutoffs in the KK
case arise at scales much larger than $1/R$, which is the typical mass scale for
the KK modes. Similarly, in the string theory example the cutoffs arise at
a scale much larger than the string scale $M_s$ which sets the mass of
the stringy modes.

In our holographic AdS setups there are various candidates that can be
responsible for alleviating the strong coupling problems. For example,
string theory in AdS has extra modes with mass of order the AdS scale,
which appear as bound states, $e.g.$ bound states of scalar
fields.\footnote{One-loop diagrams in flat space would give branch cuts,
but in AdS one finds poles that imply the presence of bound states.
These poles are visible, for instance, in the $\Gamma$-functions
of equation (78) in \cite{pkr}. They appear also as short
string states in the exact world-sheet analysis of \cite{Maldacena:2000hw}.}
Intuitively, such modes appear in the spectrum because
AdS behaves in many respects as a box. These bound states are
instrumental in providing the degrees of freedom needed to build
massive gravitons in AdS \cite{pkr}. They are expected to be responsible
for the extra interactions that soften the UV behavior and push the strong
coupling scale to the string scale.

\section{Higher dimensional CFTs and multi-gravity in AdS spaces}
\label{sec:higherD}

In the general holographic setup of \cite{kir,ack} a product of $k$ CFTs in $d$
dimensions is deformed by a set of multi-trace interactions that couple the
constituent CFTs together.\footnote{One can also consider more general
QFTs coupled together by multi-trace interactions, but we will not do so here.}
This system provides the holographic definition of a multi-gravity (or better
yet a multi-string) theory on a union of $k$ AdS$_{d+1}$ spaces. In what follows, we will
attempt to learn more about the physics of these multi-graviton theories
by studying the theory on the boundary. The new element of the analysis
that follows is a detailed study of the renormalization group equations in the space
of single-trace and multi-trace couplings. We will work in conformal
perturbation theory in the large-$N$ limit. The interpretation of the field
theory results in the bulk, combined with the observations of \cite{kir,ack},
and the lessons we draw about massive gravity in this context, will be
discussed in section \ref{sec:lessons}.

\subsection{RG flows of multi-trace couplings in conformal perturbation theory}
\label{subsec:RG}

We will focus on the most basic setup, one consisting of
a pair of conformal field theories CFT$_1$ and CFT$_2$
in $d$ dimensions with actions $\SS_1$ and $\SS_2$ respectively.
The rank of the gauge group of each theory will be denoted as $N_1$
and $N_2$. We will take a large-$N$ limit where both $N_1, N_2 \to \infty$ with
the ratio $N_2/N_1$ kept fixed. To express the common scaling
of $N_1$, $N_2$ we will set $N_1=\nu_1 N$, $N_2=\nu_2 N$
($\nu_1,\nu_2$ are fixed numbers as we send $N\to \infty$).
We want to analyze a general multi-trace deformation of the action
$\SS_1+\SS_2$ in the large-$N$ limit.

Let $\OO_1$ be a scalar single-trace gauge-invariant
operator in CFT$_1$ with scaling dimension $\Delta_1<d$ and $\OO_2$
a scalar single-trace gauge-invariant operator in CFT$_2$ with scaling dimension
$\Delta_2<d$. We normalize the operators $\OO_i \sim \tr[ \cdots ]$ so
that their two-point functions are
\beq
\label{rgnorm}
\langle \OO_i(x) \OO_i(y) \rangle=\frac{1}{|x-y|^{2\Delta_i}}~, ~~ i=1,2
~.
\eeq
The connected $n$-point functions scale with $N$ as
\beq
\label{Nscaling}
\langle \prod_{p=1}^n \OO_{i_p}(x_p) \rangle_c \sim N^{2-n}
~.
\eeq

Consider the following multi-trace deformation of the product CFT$_1\times$CFT$_2$
\beq
\label{rgaa}
\SS=\SS_1+\SS_2+\int d^d x\left[ N\sum_{i=1,2} g_i \OO_i +
\sum_{i,j=1,2} g_{ij} \OO_i \OO_j+
\frac{1}{N} \sum_{i,j,k=1,2} g_{ijk} \OO_i\OO_j \OO_k +
\cdots \right]
~,
\eeq
where the dots indicate higher multi-trace interactions. Since
$\Delta_i<d$, the single-trace couplings $g_i$ are relevant.
The double-trace couplings can be perturbatively marginal,
relevant or irrelevant depending on the sign of $d-\Delta_1$,
$d-\Delta_2$ and $d-\Delta_1-\Delta_2$. Depending on the
dimension $d$,  higher multi-trace couplings will be mostly
irrelevant, but since they can mix with each other and with
the double-trace couplings we will include them here explicitly.
The $N$-scaling of the single-trace and multi-trace operators
is such that the overall action scales like $N^2$.

We are interested in the one-loop $\beta$-functions
\beq
\label{rgbeta}
\beta_i=\frac{\d g_i}{\d t}~, ~~
\beta_{ij}=\frac{\d g_{ij}}{\d t}~, ~~
\beta_{ijk}=\frac{\d g_{ijk}}{\d t}~, ~ ~\cdots
\eeq
and the properties of the associated renormalization group flows.
In particular, we are interested to know whether there are perturbative
fixed points with non-trivial interaction between theory 1 and 2, $e.g.$ points
with $g_{12} \neq 0$. $t$ is the logarithm of the RG scale. To determine
the one-loop $\beta$-functions we will perform a standard expansion in
conformal perturbation theory keeping terms up to quadratic order in
the couplings and contributions up to the next-to-leading order in $1/N$.
To keep a concise notation with suppressed indices, it will be convenient
to denote the single-trace, double-trace, $etc.$ couplings collectively
as $g_{(1)}$, $g_{(2)}$, ....

In this approximation the RG flow equations exhibit the following structure
\begin{subequations}
\bea
\label{imrgaca}
\dot g_{(1)}&=&(d-\Delta_{(1)})g_{(1)}+C_{(1)(1)(1)} g_{(1)}^2
+C_{(1)(1)(2)} g_{(1)}g_{(2)}+
\\
&&+N^{-2}\left[
C_{(1)(1)(3)} g_{(1)}g_{(3)}
+ C_{(1)(1)(4)} g_{(1)}g_{(4)}
+C_{(1)(2)(2)} g_{(2)}^2
+C_{(1)(2)(3)} g_{(2)}g_{(3)}
\right]+\cdots
~,\nonumber
\eea
\vspace{-.5cm}
\bea
\label{imrgacb}
\dot g_{(2)}&=&(d-\Delta_{(2)})g_{(2)}+
\\
&&+C_{(2)(1)(1)} g_{(1)}^2
+C_{(2)(1)(2)} g_{(1)}g_{(2)}
+C_{(2)(1)(3)} g_{(1)}g_{(3)}
+C_{(2)(2)(2)} g_{(2)}^2+
\nonumber\\
&&+N^{-2}\left[
C_{(2)(1)(4)} g_{(1)}g_{(4)}
+C_{(2)(2)(3)} g_{(2)}g_{(3)}
+C_{(2)(2)(4)} g_{(2)}g_{(4)}
+C_{(2)(3)(3)} g_{(3)}^2
\right] +\cdots
~, \nonumber
\eea
\vspace{-.5cm}
\bea
\label{imrgacc}
\dot g_{(3)}&=&(d-\Delta_{(3)})g_{(3)}
+C_{(3)(1)(1)}g_{(1)}^2
+C_{(3)(1)(2)} g_{(1)}g_{(2)}+
\\
&&+C_{(3)(1)(3)} g_{(1)}g_{(3)}
+C_{(3)(1)(4)} g_{(1)}g_{(4)}
+C_{(3)(2)(2)} g_{(2)}^2
+C_{(3)(2)(3)} g_{(2)}g_{(3)}+
\nonumber\\
&&+N^{-2}\left[
C_{(3)(2)(4)} g_{(2)}g_{(4)}
+C_{(3)(3)(3)} g_{(3)}^2
+C_{(3)(3)(4)} g_{(3)}g_{(4)}
\right]+\cdots
~,\nonumber
\eea
\vspace{-.5cm}
\beq
\label{imrgacd}
\dot g_{(4)}=\cdots
~.
\eeq
\end{subequations}
We included terms with up to 4-trace couplings in these expressions.
The general structure is evident from the equations above.
We only add a few comments on
the notation and the basic features of these equations.

On the right hand side, the leading contribution comes from the linear terms,
which capture the classical running of the couplings. The notation is such that
in the $\beta$-function of the $n$-trace coupling $g_{i_1\cdots i_n}$ the linear
term is $(d-\Delta_{i_1 \cdots i_n})g_{i_1\cdots i_n}$ (no summation implied
here).

The quadratic terms are generated at one-loop. In this case, the notation is
such that terms of the form, say $C_{(1)(1)(2)}g_{(1)}g_{(2)}$,
are shorthand for the sum
\beq
\label{rgexamplesum}
\sum_{j,k,l=1,2}C_{i,j,kl} ~g_j g_{kl}
~.
\eeq
The constants $C_{i_1\cdots i_n,j_1 \cdots j_m, k_1 \cdots k_p}$
are proportional to the value of the three-point functions
\beq
\label{rgthreepoint}
\left\langle :\OO_{i_1}(x) \cdots \OO_{i_n}(x): ~
:\OO_{j_1}(y) \cdots \OO_{j_m}(y): ~
\int d^d z~ :\OO_{k_1}(z) \cdots \OO_{k_p}(z):
\right\rangle
\eeq
that determine the anomalous dimensions. In general, these amplitudes
receive contributions at various orders in $1/N$ both from connected
and disconnected correlation functions. Contributions that result to
wavefunction renormalization do not participate in the calculation
of the $\beta$-functions. In the RG equations \eqref{imrgaca} --
\eqref{imrgacd} we have already taken out the leading $N$ dependence
of the $C$ coefficients, which, therefore, have an $1/N$ expansion of
the form
\beq
\label{Cexpansion}
C_{(m)(n)(p)}=C^{(0)}_{(m)(n)(p)}+\frac{1}{N^2}C^{(1)}_{(m)(n)(p)}
+\cdots
~.
\eeq
$C^{(s)}_{(m)(n)(p)}$ are $N$-independent constants.

For example, $N^{-1}C_{1,1,1}$ receives its leading contribution
from the connected correlation function
$\langle \OO_1 \OO_1\OO_1\rangle_c$ and is indeed of order $1/N$.
On the other hand, $C_{11,11,11}$ receives its leading contribution at $N^0$
from the disconnected correlation function
\beq
\label{exampledisco}
\Big\langle \OO_1(x) \OO_1(y) \Big\rangle
\Big\langle \OO_1(x) \int d^dz ~\OO_1(z) \Big\rangle
\Big\langle \OO_1(y) \int d^dz ~\OO_1(z) \Big\rangle
~.
\eeq

In deriving the above RG equations, we have assumed that the single-trace
operators $\OO_1$, $\OO_2$ do not couple to other single-trace operators
in CFT$_1$, or CFT$_2$. Relaxing this assumption is not expected to change the
qualitative features of our conclusions.

\subsubsection{Fixed points at leading order in $1/N$}
\label{subsec:fixedpoints}

We first focus on the leading $\OO(1)$ terms in the $1/N$ expansion
in \eqref{imrgaca}. We  observe that by setting the
single-trace couplings to zero, no RG running is
generated and the couplings remain zero throughout the RG flow.
Hence, we can explore RG flows that evolve purely in the space of
multi-trace couplings. Setting $g_1=g_2=0$ \footnote{In this expression
$g_2$ is a single-trace coupling for theory 2 and should not be confused
with $g_{(2)}$ that refers collectively to the double-trace couplings.}
in \eqref{imrgacb} -- \eqref{imrgacd} we obtain (again, to leading order in
$1/N$)
\begin{subequations}
\bea
\label{leadimrgacb}
\dot g_{(2)}=(d-\Delta_{(2)})g_{(2)}+
C_{(2)(2)(2)} g_{(2)}^2
~,
\eea
\vspace{-1cm}
\bea
\label{leadimrgacc}
\dot g_{(3)}=(d-\Delta_{(3)})g_{(3)}
+C_{(3)(2)(2)} g_{(2)}^2
+C_{(3)(2)(3)} g_{(2)}g_{(3)}
~
\eea
\end{subequations}
with similar expressions for the higher multi-trace coupling equations.

We are looking for non-trivial perturbative fixed points of the RG equations
\eqref{leadimrgacb}, \eqref{leadimrgacc}. If we can find a solution to
the fixed point equations for the double-trace couplings $\dot g_{ij}=0$
$(i,j=1,2)$, then we can immediately satisfy the fixed point equations
for the higher multi-trace couplings. Specifically, from
the fixed point equations $\dot g_{ijk}=0$ we find
\beq
\label{rgtripleleading}
g_{ijk}=-\frac{\displaystyle\sum_{i_1,i_2,j_1,j_2=1,2} C_{ijk,i_1i_2,j_1j_2}
~g_{i_1i_2}g_{j_1j_2}}
{d-\Delta_i-\Delta_j-\Delta_k}
+\OO(g_{(2)}^3)
~.
\eeq
Similar expressions hold for the higher multi-trace couplings.

Consequently, our task is to find solutions to the fixed point equations
$\dot g_{ij}=0$. In appropriate normalization, the explicit form of these
equations is (see also appendix A of \cite{kir})
\begin{subequations}
\label{rgmaster}
\beq
\label{bdrgca}
(d-2\Delta_1)g_{11}-8g^2_{11}-8g^2_{12}=0
~,
\eeq
\vspace{-.7cm}
\beq
\label{bdrgcb}
(d-2\Delta_2)g_{22}-8g^2_{22}-8g^2_{12}=0
~,
\eeq
\vspace{-.7cm}
\beq
\label{bdrgcc}
(d-\Delta_1-\Delta_2)g_{12}-8g_{12}(g_{11}+g_{22})=0
~.
\eeq
\end{subequations}

We find the following solutions.
First, there are fixed points with no interaction between the two CFTs,
$i.e.$ fixed points with $g_{12}=0$. In this case, equations \eqref{bdrgca} --
\eqref{bdrgcc} imply
\beq
\label{rgac}
g_{11}(8g_{11}-(d-2\Delta_1))=0~, ~ ~ g_{22}(8g_{22}-(d-2\Delta_2))=0
~.
\eeq
Assuming $\Delta_1, \Delta_2 \neq d/2$, the conformal field theories
1 and 2 can be deformed individually to the non-trivial fixed
points with $g_{ii}=\frac{1}{8}(d-2\Delta_i)$. These theories are well known
\cite{witten}. One of their features is the change of the scaling dimension
of the operator $\OO_i$ from $\Delta_i$ in the undeformed theory to
$d-\Delta_i$ in the deformed. This situation is a close analog of a
corresponding matrix model example in \cite{kirniar1}.

For $\Delta_1\neq \Delta_2$ and $\Delta_1+\Delta_2 \neq d$
the fixed point equations \eqref{rgmaster} do not allow for any other solutions.
Additional fixed points can arise either when $\Delta_1=\Delta_2$ or
$\Delta_1+\Delta_2=d$ provided that $\Delta_1,\Delta_2 \neq \frac{d}{2}$.
It is enough to consider one of these cases, since the two are connected
to each other by an RG flow in a single CFT of the type mentioned in
the previous paragraph and analyzed in \cite{witten}.
We therefore consider the second case: $\Delta_1+\Delta_2=d$.

\begin{figure}[t!]
\centerline{\includegraphics[width=10cm,height=10cm]{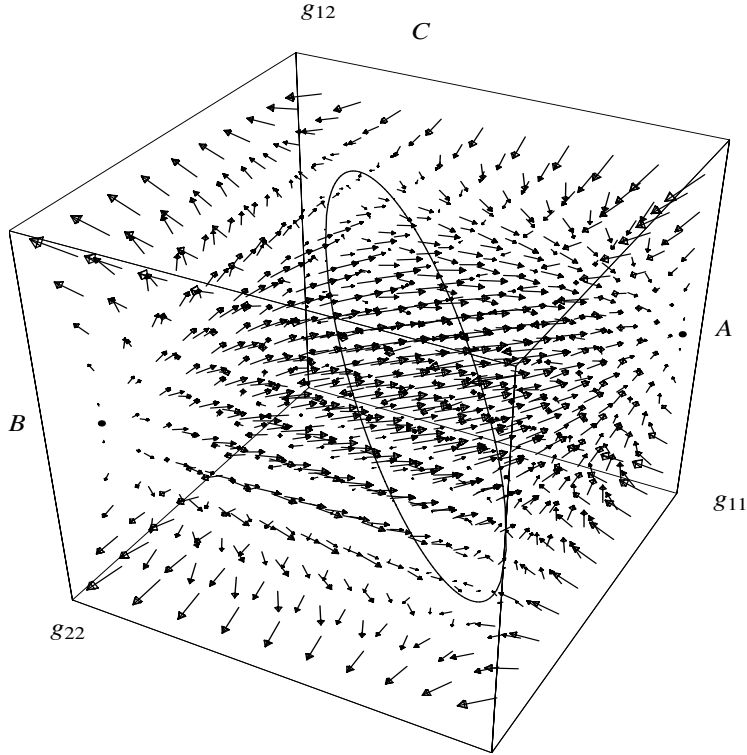}}
\caption{\small \it A circle of fixed points of the one-loop $\beta$-function in the
$(g_{11},g_{22},g_{12})$ plane for $a=\frac{d}{2}-\Delta_1=\Delta_2-\frac{d}{2}>0$.
The $\beta$-function vector field vanishes on the fixed circle $C$ at the center of
the plot, and at the special points $A=(\frac{a}{4},0,0)$ and $B=(0,-\frac{a}{4},0,0)$.
Points on the fixed circle with $g_{12}\neq0 $ and point $B$ are repellors of the
RG flow, whereas point $A$ is an attractor.}
\label{fig:gf}
\end{figure}

Here, we find a one-parameter family of new interacting fixed points
with the property
\beq
\label{rgad}
g_{11}=-g_{22}\equiv f~, ~ ~ g_{12}\equiv g~ ~ {\rm and} ~ ~
4f^2-af+4g^2=0~, ~ ~ 2a\equiv d-2\Delta_1=2\Delta_2-d
~.
\eeq
The quadratic constraint between $f$ and $g$ admits real
solutions if and only if $|g|\leq \frac{|a|}{8}$. Notice that there are no
new non-trivial solutions if $a=0$, $i.e.$ $\Delta_1=\Delta_2=\frac{d}{2}$.
Hence, assuming $a\neq 0$,  the new fixed points
lie on a circle $C$ in the $(f,-f,g)$ plane as depicted in fig. \ref{fig:gf}.
The one-loop approximation is valid for the whole circle if $|a| \ll 1$.
For generic $a$ there are always fixed points near the origin as long as
we take $|f|\ll 1$.

It is worth pointing out the striking similarity between the fixed
points \eqref{rgad} and the one-parameter family of double
scaling limits in the double-trace deformed two-matrix models
analyzed in \cite{kirniar1}. In particular, in both cases $g_{11}$ and
$g_{22}$ need to be turned on in order to get an interacting product of CFTs.

An important feature of the fixed points \eqref{rgad} is the fact that they are repellors
of the RG equations \eqref{rgmaster}. When we perturb away from the fixed circle,
there are always directions where the RG flow is repelling.
This can be verified analytically with a perturbative analysis of the RG equations
around an arbitrary point of the fixed circle (for details see appendix \ref{app:stability}).
It is hard to obtain an analytic solution of the full equations, but it is clear that as the
RG time evolves, the theory will be driven towards one of the other perturbative fixed
points that we found above or will be driven away from the origin towards a direction
where the one-loop validity of \eqref{bdrgca} -- \eqref{bdrgcc} will be lost. Indeed, it
is possible to verify this picture explicitly by analyzing the one-loop RG equations
numerically. For instance, if we focus on initial
conditions that are restricted on the $(f,-f,g)$ plane in fig.\ \ref{fig:gf} ($i.e.$ require
initially $g_{11}=-g_{22}$), we find two possible evolutions. If we start outside the fixed circle,
the RG flow takes the theory away from the perturbative region. On the other hand,
starting inside the circle takes the theory towards the perturbative fixed point
$g_{11}=\frac{d-2\Delta_1}{8}, g_{22}=0, g_{12}=0$ (this is point $A$ in fig.\
\ref{fig:gf}, assuming $\Delta_1<\frac{d}{2}$). The latter is sensible,
since this is the most attractive fixed point.

\begin{figure}[t!]
\centerline{\includegraphics[width=10.7cm,height=10.7cm]{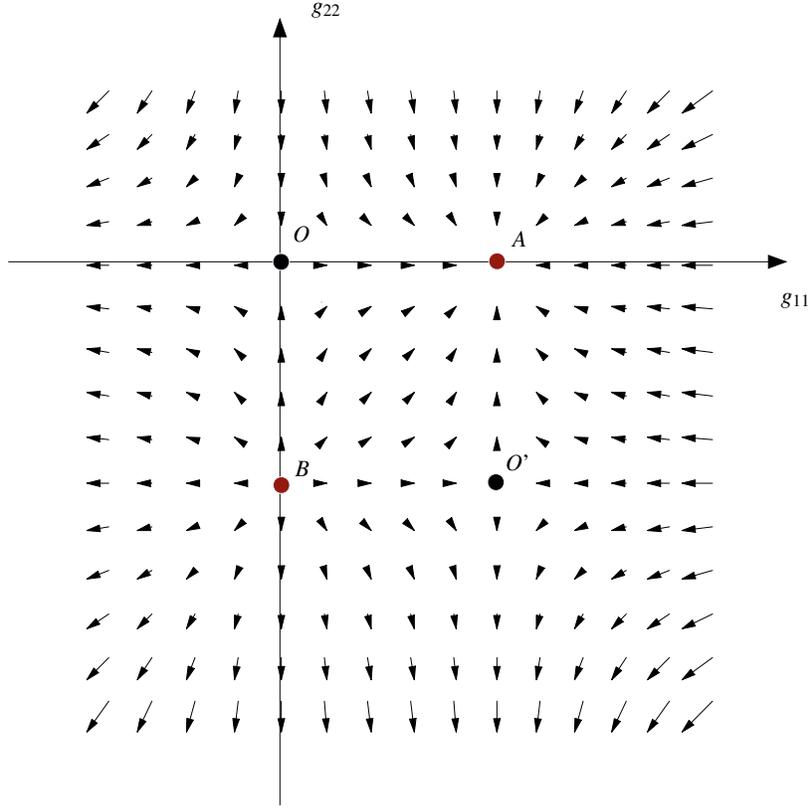}}
\caption{\small \it A 2D slice of the RG flow vector field along the $g_{12}=0$ plane.
Point $B$ is a repellor, $O$ and $O'$ are saddle points and point $A$ is an
attractor.}
\label{fig:gf2D}
\end{figure}

In fig.\ \ref{fig:gf} we also see point $B$ in the left-most corner, which is a repellor
of the renormalization group equations. At point $B$, $g_{12}=0$ and both operators
$\OO_1$, $\OO_2$ have the same scaling dimension $\Delta_1=\Delta_2<\frac{d}{2}$. Fig.\
\ref{fig:gf2D} depicts a two-dimensional slice of the $\beta$-function vector field along
the $g_{12}=0$ plane, where the $g_{12}$ component of the $\beta$-function vanishes.
Points $A$ and $B$ are the same points as in fig.\ \ref{fig:gf}. The other two special points
$O=(0,0,0)$ and $O'=(\frac{a}{4},-\frac{a}{4},0)$ are the points where the fixed circle cuts the
$g_{12}=0$ plane. As is evident from fig.\ \ref{fig:gf2D}, $O$ and $O'$ are saddle points
of the RG flow. The scaling dimensions of the operators $\OO_1$,
$\OO_2$ at these four special points are respectively the following:
at point $B$ $(\Delta_1,\Delta_1)$, at point $O$ $(\Delta_1,d-\Delta_1)$,
at point $O'$ $(d-\Delta_1,\Delta_1)$ and at point $A$ $(d-\Delta_1,d-\Delta_1)$.

\subsubsection{Fixed points at next-to-leading order in $1/N$}

Incorporating $1/N$ effects will not, in general, modify the above picture dramatically.
The most important new element is a back-reaction of the running of multi-trace
couplings to the single-trace parameters. Revisiting \eqref{imrgaca} we
discover that the next-to-leading order fixed point values of $g_i$ are
\beq
\label{gishifted}
g_i=-N^{-2}\frac{\displaystyle\sum_{j_1,j_2,k_1,k_2=1,2}
C_{i,j_1j_2,k_1k_2}~
g^{(0)}_{j_1j_2}g^{(0)}_{k_1k_2}}{d-\Delta_i}
+\OO\left(({g^{(0)}_{(2)}})^3\right)~, ~ ~ i=1,2
~,
\eeq
where $g_{(2)}^{(0)}$ denotes the leading order fixed point values
of the double-trace couplings that we found in the previous subsection.

Inserting this result back into the remaining fixed point equations $\beta_{ij}=0$,
and keeping terms up to quadratic order in $g_{ij}$, we recover the equations
\eqref{bdrgca} -- \eqref{bdrgcc} with $1/N^2$ corrections in the $C_{(2)(2)(2)}$
coefficients. The same goes for the rest of the multi-trace RG equations. The higher
order corrections can be computed similarly from the full $1/N$ expansion of
the $C$ coefficients. These will shift perturbatively the fixed circle $C$ in
fig.\ \ref{fig:gf}, but are not generically expected to spoil the basic qualitative
picture that we found in the previous subsection.

\subsubsection{Short summary}

From the above analysis in conformal perturbation theory, we conclude that there is
a line of fixed points with non-trivial double-trace interaction ($g_{12}\neq 0$) between
two CFTs, CFT$_1$ and CFT$_2$. This line exists if and only if the scaling
dimensions of the operators $\OO_1$ and $\OO_2$ are equal or sum up to the
dimension of the spacetime on the boundary. When the line exists, it is a repellor of the
renormalization group equations and a small perturbation off the fixed point locus
drives the theory towards a fixed point without interaction $(g_{12}=0)$ or
at large values of the couplings where conformal perturbation theory
breaks down. At leading order in the $1/N$ expansion, the running of the
double-trace couplings has no effect on single-trace couplings, which can be
consistently set to zero. Beyond the leading order, non-zero single-trace
couplings are generated and there are flows driving them towards finite values.

The absence of RG running for single-trace coulings at tree-level is consistent
with the bulk solution, which is unaffected and remains AdS. The bulk solution
is modified away from AdS once we take into account the subleading corrections
in $1/N$ in accordance with the non-trivial RG running induced on the boundary QFT.
On the level of the low-energy effective action in the bulk, one-loop corrections
generate extra terms that couple the two previously separate supergravity theories
and modify the original AdS$\times$AdS solution. Part of the one-loop corrections
is a quadratic mass interaction for a particular combination of the metrics.

The dual quantum gravity/string interpretation will be discussed further
in the next section. We make a short parenthesis here to present a few
concrete examples of our setup and a possible extension to multiply coupled CFTs.

\subsection{Examples}
\label{subsec:examples}

The basic ingredient of the general framework in the previous subsection
is a large-$N$ $d$-dimensional CFT that possesses a scalar single-trace
operator $\OO$ with scaling dimension $\Delta<\frac{d}{2}$.
Examples of such theories can be found in diverse dimensions.

In two dimensions the unitarity bound is $\Delta \geq 0$, so in principle
examples can be found with operators whose scaling dimensions are
arbitrarily low. An example of such a theory was presented in \cite{kir}
($c.f.$ sec.\ 7 and appendix B). It is a conformal coset CFT\footnote{This
theory can be made supersymmetric by the addition of $N^2-1$ free fermions.}
\beq
\label{2daa}
\frac{SU(N)_{k_1} \times SU(N)_{k_2}}{SU(N)_{k_1+k_2}}
~
\eeq
that can be obtained by gauging a diagonal $SU(N)$ symmetry of the WZW model
$SU(N)_{k_1}\times SU(N)_{k_2}$. The relevant parameters in this theory
are the number of colors $N$ and the 't Hooft coupling constants
\beq
\label{2dab}
\lambda_1=\frac{N}{k_1}~,~  ~ \lambda_2=\frac{N}{k_2}
~.
\eeq
In the large-$N$ limit we take $N\to \infty$ with $\lambda_i$ $(i=1,2)$ kept
fixed. This implies, in particular, that we must also scale the levels $k_i\to \infty$.

Most primary operators in this theory are in one-to-one correspondence with
triplets of representations $(R_1,R_2,R)$ with $R_1\in SU(N)_{k_1}$,
$R_2 \in SU(N)_{k_2}$, $R_1\otimes R_2 \sim R\in SU(N)_{k_1+k_2}$.
Ref.\ \cite{kir} showed that operators associated to $(R,\bar R,X)$, with
$X\in R\otimes \bar R$, are single-trace. For example, operators associated
to $(\Box,\bar \Box,1)$ are single-trace and have large-$N$ scaling dimension
\beq
\label{2dac}
\Delta_{\Box, \bar \Box,1}=\frac{\lambda_1}{1+\lambda_1}+
\frac{\lambda_2}{1+\lambda_2}
~.
\eeq
It is possible to tune $\lambda_1,\lambda_2$ so that $\Delta_{\Box, \bar\Box,1}<1$,
$e.g.$ when $\lambda_1=\lambda_2=\lambda$, we achieve this condition
when $\lambda<1$. More general primary operators in this theory are
presented in \cite{kir}.

In four dimensions, the standard example in AdS/CFT -- the $\NN=4$ SYM theory --
does not possess any scalar operators with scaling dimension $\Delta<2$. There are,
however, many $\NN=1$ super-CFTs that do have such operators. One example
is the conifold quiver theory of \cite{kwconi}. This is an $\NN=1$
$SU(N)\times SU(N)$ gauge theory with two bi-fundamental chiral multiplets
$A_i$ ($i=1,2$), and two anti-bi-fundamental chiral multiplets $B_i$. The theory flows
in the IR to a strongly coupled fixed point. The dual geometry is
$AdS_5\times T^{1,1}$. The scalar operators $\tr[A_k B_l]$ are chiral and
have scaling dimension $\Delta=\frac{3}{2}$.

Many more four dimensional examples can be found in a large class of
fixed points that arise in $\NN=1$ SQCD with zero, one or two adjoint chiral
superfields. SQCD with no adjoint chiral superfields is an $\NN=1$ gauge
theory with $N_f$ chiral multiplets $Q_i$ in the fundamental of the gauge group
(here $SU(N_c)$) and $N_f$ chiral multiplets $\tilde Q^{\tilde i}$ in the anti-fundamental.
In the large-$N$ limit, we take $N_c,N_f\to \infty$ with the ratio $x=\frac{N_c}{N_f}$
kept fixed. In the conformal window $\frac{1}{3}\leq x < \frac{2}{3}$ the theory flows
towards a non-trivial fixed point in the IR, where meson operators have scaling
dimension $\Delta=3-3x$ and therefore satisfy $1<\Delta\leq 2$.

Another example is provided by the $\NN=1$ SQCD theory with an extra chiral
multiplet $X$ in the adjoint representation of the gauge group. Let us consider
the case with a vanishing superpotential. Asymptotic freedom requires $x>\frac{1}{2}$.
This theory is believed to flow in the IR to a non-trivial fixed point. It has several types
of gauge invariant chiral operators, one class consisting of the single-trace operators
\beq
\label{4daa}
\OO^{(k)}=\tr[ X^k ]~, ~ ~ k=2,3,...
~.
\eeq
The IR scaling dimensions of these operators can be determined exactly with
the use of the $a$-maximization techniques of
\cite{Intriligator:2003jj,Kutasov:2003iy,Intriligator:2003mi}. The result is
\beq
\label{4dab}
\Delta(\OO^{(k)})=\frac{3k}{2}\frac{1-y(x)}{x}
~,
\eeq
where $y(x)$ is a known monotonically decreasing function of $x$ with
\beq
\label{4dac}
y(x\to \infty) = \frac{\sqrt 3-1}{3}
~.
\eeq
One can always arrange the free parameter $x$ so that, for any selected $k$,
$1<\Delta(\OO^{(k}))<2$.

Unfortunately, the string dual of the SQCD theories is not currently known. To
find this dual it is necessary to go beyond the supergravity approximation.

Other examples include the F-theory constructions of $AdS_5$ duals
\cite{Aharony:1998xz}, where one finds scalar operators with $\Delta=\frac{6}{5},
\frac{4}{3},\frac{3}{2}$, and the large-$N$ $(2,0)$ 3d theory dual to $M$ theory on
$AdS_4 \times S^7$, where one finds the dimension 1
\cite{Aharony:1998rm,Minwalla:1998rp}.

\subsection{Multiply coupled CFTs}

So far we focused on setups that involve a pair of CFTs coupled by multi-trace
deformations. A natural generalization involves $n$ CFTs coupled together
two-by-two by double-trace deformations with an overall action of the form
\beq
\label{multicoupaa}
S=\sum_{i=1}^n S_i+\sum_{i,j=1}^n \int d^dx~ g_{ij}\OO_i \OO_j
~.
\eeq
$\OO_i$ is a single trace operator in the $i$-th CFT, whose undeformed action
is $S_i$.

In the search of fixed points in this theory, higher order multiple trace interactions will play
the same r\^ole that they played previously in subsection \ref{subsec:fixedpoints}.
They will be sourced by the running double-trace couplings and will get shifted
to a small value near the origin at the new perturbative IR fixed points.

It is interesting to ask whether it is possible to find perturbative fixed points
at tree-level ($i.e.$ to leading order in the $1/N$ expansion), where the single-
and double-trace couplings are kept to zero, but triple-trace or higher-trace operators
are turned on. This would be a situation where three or more CFTs are coupled
in a 3- (or higher-order) junction. An example of a three-juction in $d=0$
dimensions was provided in \cite{kirniar1}, where a double-scaling limit of
three one-matrix models deformed by a triple-trace operator was found.

In higher dimensions, similar examples require triple-trace interactions that are
classically marginal. Setting $g_{(1)}=g_{(2)}=0$ to the leading $1/N$ expansion
of the RG equations \eqref{imrgaca} -- \eqref{imrgacd} we see that we can satisfy
automatically the fixed point equations $\dot g_{(1)}=\dot g_{(2)}=0$. On the right
hand side of \eqref{imrgacc} only the classical (linear) term remains, and can be
set to zero by requiring a classically marginal triple-trace deformation, $i.e.$
$\sum_{i=1}^3 \Delta_i=d$.

The unitarity bound
\beq
\label{multicoupab}
\Delta \geq \frac{d-2}{2}
\eeq
implies $\sum_{i=1}^3 \Delta_i \geq \frac{3}{2}(d-2)$.
In six dimensions we have to saturate this bound, hence the operators
are free scalars and the perturbation is an unstable cubic interaction.\footnote{We
will return to the issue of non-perturbative stability in the next section.}
We conclude that higher than binary junctions do not exist non-perturbatively
in six or more dimensions \cite{kir}. In four dimensions operators with scaling
dimension $1<\Delta<\frac{4}{3}$ satisfy the constraints. Examples of such
operators can be found in SQCD, or adjoint-SQCD. Finally, in two dimensions
the unitarity bound is $\Delta >0$ and operators with very low dimension exist.
Examples can be found in \cite{kir} (see also the previous subsection
\ref{subsec:examples}).

\section{Lessons for massive gravity}
\label{sec:lessons}

\vspace{-.2cm}
\subsection{The dual gravitational theory}

In the main example of section \ref{sec:higherD}, we considered a pair of CFTs
that has been deformed by a general multi-trace deformation.
In the tree-level approximation, the dual holographic description of this system
involves gravity (or more generally string theory) on the union of two AdS
spaces with mixed boundary conditions for the scalar fields that are dual to
the field theory operators $\OO_i$ \cite{kir,ack}. There are two massless,
non-interacting gravitons in this system at this order and the RG running of
the double-trace parameters on the boundary is encoded subtly in the mixed boundary
conditions. The RG flow on the boundary is not visible as a radial running of
the background solution away from AdS, however, it is visible in the holographic
computation of field theory correlation functions using a modified bulk/boundary
correspondence with mixed boundary conditions \cite{witten,berkooz,muck,minces,
petkou}. The presence of an unmodified AdS background in the bulk is sensible
in the absence of non-trivial renormalization of single-trace operators.

The dual string/gravity description acquires a more intricate character as soon as we start
incorporating $1/N$ effects. On the gauge theory side, we found that $1/N$ effects
shift the submanifold of fixed points and induce an RG running of single-trace
operators. On the gravity side, one-loop corrections induce a potential for the
gravitons leading to an interacting bi-gravity theory in the general spirit of Damour
and Kogan \cite{kd1}. From the point of view of the gauge theory, a diffeomorphism
breaking potential is generated, because the double-trace deformation violates
the energy-momentum conservation in each CFT individually. As a consequence of
the one-loop generated potential in gravity a linear combination of the gravitons
obtains a mass of order $g_{12}/N$. An orthogonal linear combination of the
gravitons, however, remains massless in accordance with the presence of an
unbroken diagonal diffeomorphism invariance. It is also important to stress that
it is not possible in this theory to decouple the dynamics of the massless graviton.
The two Planck scales are tied to each other.

The bulk theory in our setup is not a traditional string theory and one does not
expect the spacetime low-energy effective action in the bulk to be a local gravity
or supergravity action beyond tree-level. One can argue on general grounds, that
the worldsheet theory is non-local, hence the name non-local string theory (NLST)
\cite{nonlocal}. In our case, which involves a product of string theories,
non-local worldsheet interactions couple disconnected worldsheets
(with unequal geni in general) of the constituent string
theories. An explicit illustration of this structure to all orders in perturbation theory
was exhibited in a solvable example of coupled minimal string theories in \cite{kirniar1}.
In this toy model low-dimensional example the NLST structure appears to be
unavoidable beyond tree-level -- a lesson that seems to be generic also in
higher dimensional holographic setups such as the ones in this paper.

Some degree of non-locality is anticipated not only on the level of the worldsheet
theory but also on the level of the spacetime effective action
\cite{Aharony:2001dp,Aharony:2005sh}. An example was given in ref.\
\cite{Aharony:2001dp}, where the two-dimensional CFT dual of string theory on
$AdS_3\times S^3\times {\mathbb T}^4$ (with NSNS fluxes) is deformed by an
exactly marginal current-current deformation. The current is associated
with a $U(1)$ symmetry in the $SU(2)$ R-symmetry group of this theory.
The bulk effective action on $AdS_3$ is argued to be a Chern-Simons action
deformed by a boundary term of the form
\beq
\label{bulkactionaa}
\delta S_{bulk}= g \int_{\d AdS_3} A \wedge \tilde A
~.
\eeq
$g$ is proportional to the current-current coupling in the boundary CFT
and $A$, $\tilde A$ are the bulk Chern-Simons gauge fields. This deformation
is local as part of a three-dimensional effective action, but non-local as part
of a six-dimensional effective action on $AdS_3\times S^3$ (each of the fields
appearing in \eqref{bulkactionaa} is in a particular spherical harmonic of $S^3$).
The non-locality scale is set by the AdS radius. It was pointed out \cite{Aharony:2001dp}
that although \eqref{bulkactionaa} appears only as a boundary deformation, it has
important effects in the bulk. Further work on the spacetime non-local features of
NLST can be found in \cite{Aharony:2005sh}.

This example has a natural generalization in our context. For a pair of two-dimensional
boundary CFTs deformed by a current-current deformation of the form
\beq
\label{bulkactionab}
\delta S_{CFT}=\int d^2x  \sum_{i,j=1}^2  g_{ij}J_i(x) \tilde J_j(\bar x)
\eeq
the bulk deformation is
\beq
\label{bulkactionac}
\delta S_{bulk}=\int _{\d AdS_3} g_{ij} A_i \wedge \tilde A_j
~.
\eeq
In six dimensions this action has a non-local piece with a double integration
over both universes: one integration running over the first $S^3$ and the other
over the second.

The most interesting features of the bulk theory arise as quantum
corrections to the classical action.\footnote{Notice that we consider quantum
corrections that involve matter fields running in the loops, but no gravitons.
This is consistent as long as the matter field theory cutoff is hierarchically smaller
than the Planck scale. Hence, gravity remains semiclassical in our discussion.}
For example, it is in such corrections, where one sees more dramatically in the
action the coupling between the two AdS theories with new explicit bulk interactions,
$e.g.$ it is in these corrections where a quadratic mass term for a linear combination
of the two gravitons appears giving rise to a non-trivial theory of massive gravity.
As we said, the resulting gravitational effective action is expected to be an action in
the general spirit of non-linear multi-gravity actions proposed in \cite{kd1}.
Determining the precise form of this action in our case appears to be a daunting
task. In the rest of this section, we will try to diagnose some of its properties
from the boundary non-gravitational theory.

The structure of the RG flow equations on the boundary makes a definite prediction
for what type of solutions we should anticipate from the equations of motion of
the loop corrected effective action in the bulk. On general grounds, we anticipate
fixed points of the RG flow equations on the boundary QFT to correspond to
AdS$\times$AdS solutions in the bulk. RG flows interpolating between fixed points
correspond to domain wall solutions interpolating between different AdS$\times$AdS
solutions.

Hence, the existence of the non-trivial fixed points in section \ref{subsec:RG},
predict that by fine-tuning the double-trace parameters, we should be able to find
AdS$\times$AdS solutions with two interacting gravitons, a linear combination of
which is massive. Since $g_{12}\neq 0$, the theories in each AdS interact with each
other in a non-trivial manner according to the general rules of NLSTs. In these
solutions the scalar fields dual to the single-trace operators $\OO_i$ acquire a
non-trivial profile. In gauge theory, we find a corresponding non-zero value for
the single-trace couplings $g_i$ (see eq.\ \eqref{gishifted}).
From the point of view of the bulk effective action, the excited scalar fields are
needed to counterbalance the extra terms in the equations of motion from the
one-loop generated corrections.

For generic values of the double-trace parameters, outside the special submanifold
of fixed points, there is a non-trivial boundary RG flow. This flow translates in the
bulk to a background that is asymptotically AdS$\times$AdS, but
deviates from AdS as we move away from the conformal boundary. Depending
on the bare values of the double-trace parameters two types of RG flows are possible
in gauge theory. One in which the RG flow drives the theory to strong coupling
outside the range of validity of conformal perturbation theory, and another one in
which the RG flow drives the theory back to a perturbative fixed
point with $g_{12}=0$, $i.e.$ zero interaction between the two CFTs. We cannot
say much about the first case, unless we go beyond the perturbative analysis of
section \ref{sec:higherD}. For the second case, however, we expect to have a bulk
solution that interpolates radially between two AdS$\times$AdS solutions. In the UV part
of the geometry gravity is massive and there is non-trivial interaction between the
two AdS universes. The leading $1/N$ value of the mass of the gravitons
is controlled by the bare value of the double-trace parameter $g_{12}$ and can be
determined with a one-loop computation in the tree-level theory as in \cite{kir,ack}.
In the IR part of the geometry, one recovers a patch of spacetime
with two massless, non-interacting gravitons. This tentative class of domain wall
solutions provides a dynamical mechanism of switching off the mass of gravitons. No
vDVZ discontinuity is anticipated to prevent the existence of such solutions in this
case. The absence of the vDVZ discontinuity is a well known fact for AdS backgrounds
\cite{porrativvd}.

As a general observation, we point out the fact that  fine-tuning is necessary
in order to arrange for a solution where gravity is massive everywhere in spacetime.
For generic boundary data, the solution can develop patches of spacetime where
the one-loop generated mass of the gravitons is dynamically switched off.
In that sense,  {\it massive gravity is not generic}.

\subsection{Instabilities and strong coupling}
\label{instastrong}

Massive graviton theories are in general unstable \cite{deser,insta}. The
unstable mode is a ghost ($i.e.$ a mode with a wrong sign kinetic term),
and sometimes also a tachyon (a mode with a negative mass squared). The
source of the instability lies in the non-linear structure of the kinetic terms.
For instance, a quadratic Lagrangian in four spacetime dimensions that
comprises of $(i)$ the Einstein-Hilbert action, expanded to leading order around
flat space, and $(ii)$ a quadratic term for the metric perturbation of the
Fierz-Pauli form $h_{\mu\nu}h^{\mu\nu}-(h_\mu^\mu)^2$ is an action that describes
correctly at the linearized level a massive two-tensor field with five propagating
degrees of freedom. The problem arises when we try to expand around a
general background. Taking into account higher order non-linear terms,
$e.g.$ the higher order terms in the expansion of the Einstein-Hilbert action,
one finds \cite{deser} an additional propagating scalar mode that has a wrong
sign kinetic term. This unstable mode, commonly known as the Boulware-Deser
mode, is a generic feature of Lorentz invariant massive gravity independent
of the non-linear extension of the theory and the background.

The examples of massive gravity on AdS$\times$AdS backgrounds
in this paper are described by an effective action, which is not expected,
as we said, to be a conventional local gravity or supergravity action.
It is natural to ask whether we see any signs of an instability (ghosts or
tachyons) in this theory. Since we do not have an explicit formulation of
this action, it is advisable to look for any signs of instability, or pathology,
in the dual gauge theory description.

In conformal perturbation theory (perturbation theory in $g_{(2)}$), the dual
gauge theory appears to be a
well-defined CFT, provided that we tune a number of  single-trace and
multi-trace couplings to a special set of values. The non-perturbative (in $g_{(2)}$)
consistency of this theory is less clear. A potential worry is that the multi-trace
deformation will create a de-stabilizing potential for the fundamental fields
of the constituent CFTs 1 and 2.

For example, if we couple two copies of the conifold quiver $\NN=1$ SCFT
and perturb by multi-trace interactions that involve the scalar operators
$\tr[ A_k B_l]$, we will necessarily get a potential which is unbounded from
below for the bi-fundamental fields $A_k$, $B_l$. In similar examples
constructed out of the one-adjoint SQCD theory, deformations involving the
operators $\OO^{(2k)}$ (see eq.\ \eqref{4daa}) are even under the parity
$X\to -X$, hence it is not immediately obvious whether a de-stabilizing
direction occurs in field space. To see what happens, one has to determine
the non-vanishing single-trace and multi-trace couplings of the deformation.
At the fixed point circle of section \ref{sec:higherD}, an infinite number of
multi-trace couplings is turned on already at tree level. In the large $N$ limit,
however, the leading order potential for a single eigenvalue of the adjoint
field $X$ comes from the double-trace deformation. Since $g_{11}=-g_{22}$
for any of the fixed points \eqref{rgad}, we conclude that there is always an
unstable direction for the $X$ eigenvalues of either theory 1 or 2.

These observations seem to imply that a non-perturbative instability is a
generic feature of the models in this paper, at least when the multi-trace
deformations break the spacetime supersymmetry.\footnote{Multi-trace
deformations that preserve some amount of supersymmetry might be
better candidates for a non-perturbatively well-defined model. Double-trace
deformations that preserve $\NN=1$ supersymmetry must be the upper
components of an $\NN=1$ supermultiplet. Hence, for such deformations
to be classically relevant we must use single-trace operators with scaling
dimensions $\Delta$ in the range $\frac{d-2}{2}<\Delta<\frac{d-1}{2}$.
In $d=4$, examples of such operators can be found in SQCD.}

We believe that this type of instability is distinct from the (potential)
existence of ghosts in the bulk effective theory. First, the non-perturbative
instability in gauge theory appears also in multiple-trace deformations of
a single CFT, where the dual gravitational theory has no massive graviton.
Such a case was analyzed, for example, in \cite{Hertog:2004rz}, where a
three dimensional CFT is deformed by a marginal cubic power of a
single-trace operator. Moreover, it has been argued in concrete examples,
\cite{insta} that the Boulware-Deser (ghost) instability in massive gravity is
intimately related to the strong coupling problems of massive gravity. We
will argue in a moment that there is no strong coupling problem
in our holographic setups. This suggests that there are no
Boulware-Deser instabilities in the theories we study.

The other generic feature of massive graviton theories is, as we said, the
presence of a scale where the scalar mode of the graviton becomes strongly
coupled and the effective action requires a UV completion. Our estimates in
section \ref{sec:estimates} indicated that this scale is set in our examples by
the AdS scale. However, we pointed out that these estimates can very well
be a mirage in a theory with extra fields and extra interactions beyond the
self-interactions of the gravitons.

Indeed, the theory has new degrees of freedom starting at the AdS scale, namely
the bound states of the scalars. These provide the degrees of freedom that make
one of the gravitons massive. There is no signal of strong coupling both at the
boundary QFT and the bulk theory, and we still expect the gravitational description
to break down at the string scale, where it will be completed by the string theory
description. We therefore deduce that the bound states are instrumental in
alleviating the strong coupling problem of the massive graviton by providing
softening interactions that push the strong coupling scale of the (bi)-gravity
theory to the string scale.

\section{Discussion}
\label{discussion}

\vspace{-0.2cm}
\subsection{Summary}

In this paper we considered a holographic context that realizes interesting
theories of massive gravity. On the field theory side, a product of conformal
field theories is deformed by a multiple trace deformation that couples together
different CFTs in this product. Analyzing the $\beta$-functions of single-trace
and multi-trace operators in conformal perturbation theory for a simple case
of a pair of CFTs, we found a one-parameter family of non-trivial fixed
points with non-vanishing interaction. Three of the most characteristic features
of these fixed points are:
\begin{itemize}
\item[$(i)$] The fixed points exist only when the single-trace operators participating
in the deformation have equal scaling dimensions, or dimensions adding up
to the spacetime dimension.
\item[$(ii)$] An infinite number of multi-trace couplings is generated by loop
corrections. Single-trace couplings are also generated beyond the leading
$1/N$ order.
\item[$(iii)$] The fixed points are repellors of the RG equations and perturbing
away from them drives the theory towards a fixed point where the coupling
between the CFTs disappears, or towards strong coupling outside the range
of validity of conformal perturbation theory.
\end{itemize}

The dual gravitational description of this system is a non-local multi-string
theory on a union of AdS spaces (or more generally on a domain wall solution
interpolating between unions of AdS spaces). The low-energy effective action
is a multi-graviton theory generalizing the paradigm of multi-gravity actions
proposed in \cite{kd1}. A non-trivial potential for the gravitons is generated
here by quantum corrections where gravity stays semiclassical, but matter fields
are allowed to run in the loops. The leading term in this potential is always a
quadratic mass term for a linear combination of gravitons. The field theory description
suggests that these non-trivial theories of massive gravity are not plagued by the usual
problems.

A naive estimate based on the self-interactions of a massive graviton in AdS implies
that the vector and scalar modes become strongly coupled at a scale set by the AdS
scale. The gauge theory description suggests that this estimate is too strong.
There are bound states in the bulk at the AdS scale that signal both a non-locality of the
bulk theory at that scale as well as provide the relevant degrees of freedom that
alleviate the strong coupling problem and soften the UV behavior of the massive
graviton amplitudes. It would interesting to study this mechanism in detail.

From this study the following conclusions can be abstracted for the gravitational side:
\begin{itemize}
\item If the diffeomorphism invariance breaking is ``small'' (here ${\cal O}(1/N)$), then
there are no strong coupling problems and no Boulware-Deser instabilities.

\item The existence of stable massive graviton backgrounds requires fine tuning.

\item Without fine tuning the spacetime background backreacts in order to erase the
effects of the graviton mass.
\end{itemize}

\subsection{Other interesting aspects of our work}
\label{extensions}

We conclude with a few additional remarks on some
other interesting aspects of this work.

\subsubsection{Multi-verse vs multi-throat}
\label{multithroat}

Effective theories with two gravitons have been discussed in seemingly different
contexts. These include multi-throat CY compactifications, \cite{multithroat} which
are very popular in string cosmology, as well as Randall-Sundrum (RS) setups,
where the two sides of the Planck brane are not related by symmetry but are
independent parts of the five-dimensional spacetime \cite{padilla}. This was
further generalized to several spaces meeting at the UV RS brane, \cite {grosj}.
Both of these realizations are cutoff versions of a single UV theory that factorizes
in the IR.

A simple prototype model of such multi-throat solutions in the AdS/CFT correspondence
is provided by the Higgs branch of the large-$N$ $\NN=4$ $U(2N)$ gauge theory in four
dimensions. At strong 't Hooft coupling this theory is described by a string theory/gravity
living on AdS$_5$.\footnote{We may neglect for this discussion the extra five sphere.}
This theory has obviously a single graviton. Now consider turning on a scalar vev that
breaks the gauge group to $U(N)\times U(N)$ and gives a mass $M$ to the bi-fundamental
vector multiplets made up from the massive W-bosons of the broken part of the original gauge
group.

At energies $E\gg M$ the theory can be described in terms of a single graviton
on an asymptotically AdS background. On the other hand, at energies $E\ll M$
a better description is given in terms of two throats \cite{kw} and two distinct AdS geometries
with two effective low-energy gravitons. From the field theory point of view, in this regime
we may integrate out the massive multiplets to generate an infinite sequence of double-trace
interactions that couple the two effective gauge theories. Therefore, in this regime it is not
possible to distinguish the qualitative physics of this theory and the ones we are describing
in this paper. Both are described as bigravity theories with a potential that gives a mass to
one of the two gravitons (although here the mass will be exponentially suppressed).

The question we want to pose is this: is there a fundamental
difference between such theories (namely theories coupled by
bi-fundamental fields) and the theories we describe in this paper (coupled by relevant
or marginal multiple trace operators)? If the double trace operators that we perturb by
in the UV are relevant then we can exclude that they can be obtained by decoupling
some massive bi-fundamental fields. The marginal case is more subtle and it is in
principle possible that such a case could be obtained in the decoupling limit.
We cannot exclude that some of our examples could be potentially viewed as subtly
interacting string theories after the messengers between the two worlds have (almost)
decoupled.

It is also interesting to note that viewed both from the QFT point of view or the gravitational
point of view, the generic expectation for a low energy theory is to be maximally ``fragmented'':
a collection of many ``hidden sectors'', that communicate only at a higher energy, and the
extreme case of the theories described in this paper, only at infinite energy. This observation
suggests a different picture for a ``landscape'' of worlds, each associated with a throat or a
separate low energy QFT, communicating via high energy. A cosmological exploration of
such an idea seems interesting.

\subsubsection{Lorentz-violating theories in a holographic context}
\label{lorentzviolation}

In the setups described in this paper the one-loop generated potential for the
gravitons respects the Lorentz invariance of the system (which is part of the AdS
invariance here). One could also envisage the case of non-invariant terms. There
are several ways to obtain such terms in a holographic context. A general setup that
would generalize the one in this paper would be to add lower-dimensional defects
in the CFTs along the lines of \cite{kra,bach}. A special case of this construction is
related to the proposal in \cite{hc}\footnote{See also section 7.8 of \cite{review}.}:
that 4d gravitons are glueballs of a 4d gauge theory, whose dual is a 5d string
theory. As shown in \cite{kn} such gravitons must be necessarily massive.

\subsubsection{Designer multi-gravity, cosmological applications}

One of the main motivations behind this work is cosmology and the
potential for interesting time-dependent solutions in massive (multi)gravity.
The cosmological constant problem is one of the problems one would
like to address in this context.

Within our framework, a potentially interesting direction is to analyze the
multi-trace deformations in this paper in the presence of non-vanishing
$\OO(1)$ single-trace couplings. In the bulk, this would be a multi-gravity
generalization of designer gravity \cite{Hertog:2004ns}. AdS cosmologies
in a single AdS have been discussed in this context by the reference
\cite{Hertog:2005hu}.

Concerning direct applications to cosmology, the situation is more obscure.
The regime of graviton masses studied here does not allow a direct application inside AdS$_4$.
It leaves however open the possibility of embedding a 4d nearly flat universe inside a higher
dimensional AdS space while keeping the IR screening of gravity. An assessment of the phenomenological viability
of such a possibility is interesting.

\subsubsection{Multiply coupled CFTs and deconstruction of gravitational dimensions}

In view of our results we may revisit theories that deconstruct gravity \cite{deco,kd1,AHSCH}.
In our language, to reconstruct a KK circle we consider $\hat N$ copies of a given
$(d-1)$-dimensional CFT, CFT$_i$, and we couple CFT$_i$ with CFT$_{i+1}$ via a
double trace perturbation $h_{i,i+i}\int {\cal O}_i{\cal O}_{i+1}$. We also couple the
$\hat N$-th theory with the first to create a topological circle. In the case where all couplings
are identical we expect to reconstruct a uniform circle. In this construction, there will
be a single massless graviton and $\hat N-1$ massive gravitons with masses
$m\sim {h\over N\ell}$. The $(d+1)$-dimensional manifold is $AdS_{d}\times S^1$.
The radius $R$ and $(d+1)$-dimensional Planck scale can be written in terms of
the $d$-dimensional Planck scale as,  \cite{AHSCH}
\be
R={\hat N\over m}\sp (M_{d+1})^{d-1}=(M_d)^{d-2} ~m
\ee
Using the various estimates we compute (for adjoint large-$N$ CFTs)
\be
{R\over \ell}\sim {N\hat N\over h}\geq 1\sp (M_{d+1}\ell)^{d-1}\sim h~N\gg 1
\sp {M_d\over M_{d+1}}\sim h^{1\over d-1}~N^{d\over (d-1)(d-2)}\gg 1
\ee
which establish the hierarchy
\be
{1\over R}\ll {1\over \ell}\ll M_{d+1}\ll M_d
\ee
Although the naive cutoff appears to be at the AdS scale the real strong couplings seem
to appear at $M_{d+1}$. This is similar to the case of usual KK compactification with
the difference being that now $M_d$ is the fundamental scale and $M_{d+1}$ is the
derived scale. Considering products of CFTs coupled by marginally relevant couplings
will in general generate higher-dimensional manifolds that do not have a product
structure. This direction deserves further study in view of the interest to generate a
quantum gravitational theory in four or higher dimensions in terms of renormalizable
gravity in three dimensions.

\medskip
\section*{Acknowledgements}
\noindent

We would like to thank Igor Klebanov and  Ioannis Papadimitriou for discussions.
We are especially indebted to Massimo Porrati for sharing his knowledge on many
of the issues described here and for providing enlightening suggestions. EK thanks
the Galileo Galilei Institute for Theoretical Physics for the hospitality and the INFN for
partial support during the completion of this work. This work was partially supported
by  ANR grant, ANR-05-BLAN-0079-02, RTN contracts MRTN-CT-2004-005104 and
MRTN-CT-2004-503369, CNRS PICS \#~  3059, 3747, and 4172 and by a European
Union Excellence Grant, MEXT-CT-2003-509661.

\section*{Appendices}

\begin{appendix}

\section{Estimates of strong coupling scales for massive gravity in $AdS_d$}
\label{apa}

In this appendix we consider a massive gravity theory in an (asymptotically) AdS$_d$
background. We will provide an estimate of the scales where self-interactions of the
graviton become strong using the formalism of \cite{ags}. For that purpose we will
introduce St\"uckelberg fields in order to keep a better track of the behavior of the
scalar and vector modes of the graviton which are propagating degrees of freedom in
the presence of a graviton mass term. In this section all numerical coefficients will be set to unity
since we are interested in order of magnitude estimates. We will ignore the tensor
structure of the graviton as it is irrelevant for our purposes.

The basic lowest-derivative action in $d$ dimensions is the Einstein-Hilbert action
plus a general potential for the graviton
\be
S=S_0+S_V\sp S_0=(M_d)^{d-2}\int d^dx \sqrt{g} ~
\left(R+{d(d-1)\over \ell^2}\right)\sp S_V=(M_d)^{d-2}~m^2~V(h)
~.
\label{m46}
\ee
$m$ is the graviton mass and
$h$ the graviton perturbation around a background value $g^{(0)}$, $i.e.$ $g=g^{(0)}+h$.
For us $g^{(0)}$ is the metric of AdS. Introducing the Goldstone modes as in \cite{ags}
by shifting $h\to h+\d A+\d^2\phi$ we obtain kinetic terms for $\phi$ and $A$ of the form
\be
S_{kin}\sim \int Z^2~(\d \phi)^2 +(M_d)^{d-2}m^2(\d A)^2\sp Z\sim
{m(M_d)^{d-2\over 2}\over \ell}\sqrt{1+m^2\ell^2}\sim (mass)^{d+2\over 2}
~.
\label{m47}
\ee
$\ell$ is the AdS scale of the background metric.
The canonically normalized fields are
\be
\phi_c=Z\, \phi\sp A_c=m\, (M_d)^{d-2\over 2}\, A
~.
\label{m48}
\ee

The interactions of the Goldstone modes emanating from the graviton potential are of
the general form
\be
I_{p,q}\sim m^2 (M_d)^{d-2} ~(\d A)^p~(\d^2\phi)^q \sim
{(m^2 (M_d)^{d-2})^{1-{p\over 2}}\over Z^q }
~(\d A_c)^p~(\d^2\phi_c)^q
\label{m49}
\ee
where $p,q\geq 0$ are integers. In particular, when $p=0$, $q\geq 3$ and when
$q=0$, $p\geq 2$.

An interaction of the form $\Lambda^{-n}\int \OO$, where $\OO$ is a composite of
normalized fields, becomes strong at energies $E\sim \Lambda$. Therefore,
from an effective field theory point of view $\Lambda$ is the UV cutoff scale
associated with this interaction. From (\ref{m49}) we obtain the following UV
cutoff scales
\be
\Lambda_{p,q}\sim {\left(m~(M_d)^{d-2\over 2}\right)^{2(p+q-2)\over
d(p+q-2)+2q}(1+m^2\ell^2)^{q\over d(p+q-2)+2q}\over \ell^{2q\over d(p+q-2)+2q}}
~.
\label{m50}
\ee

The following two parameter regimes are of particular interest:
\begin{itemize}
\item[$(i)$] {\it The ``flat space regime''}: $m\ell\gg 1$.
In this case, the mass gap lies in the region where the AdS curvature is invisible.
For energies $E\gg 1/\ell$ the massive gravity theory lives effectively in flat spacetime.
The kinetic factor is $Z\sim m^2M^{d-2\over 2}$.
\item[$(ii)$] {\it The ``AdS regime''}: $m\ell\ll 1$.
In this case, the mass gap is in the AdS region where the AdS curvature is visible.
For energies $m\ll E\ll {1\over \ell}$ the massive graviton theory resides in a space
with a visible curvature, whereas in the regime $E\gg {1\over \ell}$ we are effectively
in flat space. The kinetic factor is $Z\sim {mM^{d-2\over 2}\over \ell}$.
\end{itemize}

\subsection{Cutoffs in the flat space regime}

Taking the limit $m\ell\gg 1$ in (\ref{m50}) we obtain
\be
\Lambda_{p,q}\sim M_d~\left({m\over M_d}\right) ^{2(p+2q-2)\over
d(p+q-2)+2q}
~.
\ee
The lowest cutoff, when $m\ll M_d$, arises when the exponent
${2(p+2q-2)\over d(p+q-2)+2q}$ is maximized. This occurs when $p=0,q=3$ with
\be
\Lambda_{0,3}\sim M_d~\left({m\over M_d}\right) ^{8\over d+6}
~.
\label{m57}
\ee
If the leading $p=0$ terms are tuned to zero in the potential, then the next
smallest cutoff appears at  $q\to \infty$ and/or $p=2,q=1$
\be
\Lambda_{p,\infty}\sim \Lambda_{2,1}\sim M_d~\left({m\over M_d}\right) ^{4\over d+2}
~.
\label{m58}
\ee

For $d=4$ these estimates are in agreement with \cite{ags}, with
$ \Lambda_{0,3}=\Lambda_V$ and $ \Lambda_{2,1}=\Lambda_{AGS}$.
For general $d$
\be
{\Lambda_{0,3}\over \Lambda_{2,1}}\sim \left({m\over M_d}\right) ^{4(d-2)\over (d+2)(d+6)}
\label{m67}
\ee
indicating that $\Lambda_{2,1}$ is always the larger cutoff above two dimensions.
Finally, we observe some cases where the phenomenologically interesting cutoff
$\Lambda_*$ appears: in $d=10$, $\Lambda_{0,3}=\Lambda_{*}\sim \sqrt{mM_d}$,
whereas in $d=6$,  $\Lambda_{2,1}=\Lambda_{*}\sim \sqrt{mM_d}$.

\subsection{Cutoffs in the AdS regime}

In the AdS regime $m\ell\ll 1$, hence from (\ref{m50}) we obtain
\be
\Lambda_{p,q}\sim {\left(mM_d^{d-2\over 2}\right)^{2(p+q-2)\over
d(p+q-2)+2q}~\ell^{-{2q\over d(p+q-2)+2q}}}
~.
\label{m59}
\ee
Remember that the massive graviton theories of interest in this paper are necessarily
in this regime, because, as we saw in section \ref{gm}, the combination $m\ell\ll 1$ is
always suppressed by a positive power of $N$.

To proceed further we need to relate the scales $m, M_d$ and $\ell$. Their relation
depends on the type of large-$N$ theory we consider within the classification
of section \ref{estimate}.

\begin{itemize}

\item[(1)] {\it Standard adjoint U(N) gauge theories}.
In this case
\be
m\sim {h\over N\ell}\sp M_d^{d-2\over 2}\sim {N\over \ell^{d-2\over 2}}
\sp m(M_d)^{d-2\over 2}\sim {h\over \ell^{d\over 2}}
\label{m60}
\ee

Notice that all $\Lambda_{p,q}$ are of the same order for $h\sim \OO(1)$,
\be
\Lambda_{p,q}\sim {h^{2(p+q-2)\over
 d(p+q-2)+2q}\over \ell}\sim h^{-{4q\over d(d(p+q-2)+2q))}}~
 \left(m~M_d^{d-2\over 2}\right)^{2\over d}
\label{m61} \ee
As $h$ can be much smaller than one, the minimum of the exponent above is
of interest. This minimum is equal to $-2/d$ and we therefore have
\be
\Lambda_{min}\sim \Lambda_{0,2}\sim \Lambda_{p,\infty}\sim {1\over \ell}\sim
{1\over h^{2\over d}} \left(m~M_d^{d-2\over 2}\right)^{2\over d}
~.
\ee

According to this estimate the low-energy effective description is unreliable at
distances shorter than the AdS scale. For $d=4$, in particular, we obtain
\be
\Lambda_{min}\sim {1\over \ell}\sim {1\over \sqrt{h}}\sqrt{m~M_4}
\sp m\ell\sim \sqrt{h}\sqrt{m\over M_4}
\label{m62}
\ee
so when $h\ll 1$ the cutoff is hierarchically larger than $\Lambda_*\sim  \sqrt{m~M_4}$.

\item[(2)] {\it M2-type large-$N$ CFTs}.
Using
\be
m\ell\sim {h\over N^{3\over 4}}\sp M_d\ell\sim N^{3\over 2(d-2)}
\label{m64} \ee
and (\ref{m59}) we obtain
\be
\Lambda_{min}\sim \Lambda_{0,2}\sim \Lambda_{p,\infty}\sim {1\over \ell}
\label{m63}
~.
\ee

For $d=4$ we recover a formula similar to (\ref{m62}).

\item[(3)] {\it M5-type large-$N$ CFTs}.
Using
\be
m\ell\sim {h\over N^{3\over 2}}\sp M_d\ell\sim N^{3\over (d-2)}
\label{m65} \ee
and (\ref{m59}) we obtain again
\be
\Lambda_{min}\sim \Lambda_{0,2}\sim \Lambda_{p,\infty}\sim {1\over \ell}
\label{m66} \ee
and a formula similar to (\ref{m62}) for $d=4$.

\end{itemize}

We conclude that in all of the above holographic cases our estimate for the lowest
UV cutoff of the effective low energy (massive) gravity is the AdS scale itself.
In the main text, we argue that this cutoff is in fact a mirage and that it is important
to take into account all the interactions, not just the self-interactions of the graviton
modes.

\section{Two instructive examples of UV cutoffs}
\label{instructive}

\vspace{-.2cm}
\subsection{KK gravitons}
\label{app:KKgravitons}

Starting from a gravity theory in $D$ dimensions with Planck scale $M_D$,
we can compactify $D-d$ dimensions on a torus, which for simplicity we will
take to have a uniform radius $R$ much longer than the fundamental Planck length,
$i.e.$ $M_DR\gg 1$. Then, the lighest KK graviton has mass $1/R$ and we
have the scales
\be
m\sim {1\over R} \sp M_d^{d-2}=R^{D-d}~M_D^{D-2}\sp {M_d\over M_D}
\sim (M_DR)^{D-d\over d-2}\gg 1
~.
\label{m72}\ee
The last relation indicates that the lower dimensional Planck scale is much larger
than the higher dimensional one, and the theory decompactifies before the
$d$-dimensional gravity becomes strong.

We can now estimate the two main cutoffs in the $d$-dimensional theory
from (\ref{m57}) and (\ref{m58})
\be
\Lambda_{0,3}\sim {1\over R}(M_D~R)^{D-2\over d+6}\sp \Lambda_{2,1}
\sim {1\over R}(M_D~R)^{D-2\over d+2}
~.
\label{m73}
\ee
Consequently,
\be
R\Lambda_{0,3}\sim (M_DR)^{D-2\over d+6}\gg 1~,~ ~{\Lambda_{0,3}\over M_d}
\sim (RM_D)^{-{8(D-2)\over (d-2)(d+6)}}\ll 1
~ , ~~ {\Lambda_{0,3}\over M_D}\sim (RM_D)^{D-d-8\over d+6}
~,
\label{m74}
\ee
\be
R\Lambda_{2,1}\sim (M_D~R)^{D-2\over d+2}\gg 1~,~~
{\Lambda_{2,1}\over M_d}\sim (RM_D)^{-{4(D-2)\over (d-2)(d+2)}}\ll 1
~,~~  {\Lambda_{2,1}\over M_D}\sim (RM_D)^{D-d-4\over d+2}
~.
\label{m75}\ee
Assuming $D-d>8$ we obtain the hierarchy
\be
{1\over R}\ll M_D\ll \Lambda_{0,3}\ll M_d
~,
\label{m76}\ee
while in the opposite case $D-d<8$
\be
{1\over R}\ll  \Lambda_{0,3}\ll M_D\ll M_d
~.
\label{m77}\ee
Similarly for $D-d>4$ we obtain
\be
{1\over R}\ll M_D\ll \Lambda_{2,1}\ll M_d
\label{m78}\ee
and in the opposite case $D-d<4$
\be
{1\over R}\ll  \Lambda_{2,1}\ll M_D\ll M_d
~.
\label{m79}\ee
In both cases where the cutoffs are below $M_D$ we are forced to conclude that
they are ineffective since the theory is well defined in this regime with a $D$-dimensional
massless graviton, and the strong couplings cancel out as it was explicitly shown in
\cite{sch}.

\subsection{Stringy Massive gravitons}
\label{app:stringgravitons}

String theories contain an infinite number of massive spin-two states appearing
in the oscillator spectrum. In closed superstring theory the first such state is
$b^{\mu}_{-{1\over 2}}a^{\rho}_{-1}\bar b^{\nu}_{-{1\over 2}}\bar a^{\rho}_{-1}|p\rangle$
with a mass $m$ of order $M_s$. Therefore, in ten dimensional perturbative string
theory we have
\be
m\sim M_s\sp M_{10}^8\sim {M_s^8\over g_s^2}\sp g_s\ll 1
~.
\label{m68}
\ee
Applying the cutoffs (\ref{m57}) and (\ref{m58}) in the flat space regime we find
\be
\Lambda_{0,3}\sim \sqrt{mM_{10}}\sim  g_s^{-{1\over 8}}M_s\gg M_s\sp
{\Lambda_{0,3}\over M_{10}}=g_s^{1\over 8}\ll 1
~,
\label{m69}
\ee
\be
\Lambda_{2,1}\sim \left(mM_{10}^2\right)^{1\over 3}\sim  g_s^{-{1\over 6}}M_s\gg M_s
\sp {\Lambda_{2,1}\over M_{10}}=g_s^{1\over 12}\ll 1
~.
\label{m70}
\ee
We conclude that the theory exhibits the following hierarchy of scales
in the perturbative regime $g_s\ll 1$ around flat space
\be
m\sim M_s \ll \Lambda_{0,3}\ll \Lambda_{2,1}\ll M_{10}
~.
\label{m71}
\ee

We know, however, that the string theory perturbation theory is well defined at any
scale hierarchically below the Planck scale, and therefore we deduce that
the cutoffs  $\Lambda_{0,3}$ and  $\Lambda_{2,1}$ are merely mirage effects.
There is no associated strong coupling problem in this system, most probably
via judicious cancelations as those observed in the KK case \cite{sch}.

\section{RG flow stability near the fixed circle}
\label{app:stability}

In this appendix we demonstrate the claim in section \ref{subsec:RG}
that the fixed points on the circle $C$ of fig.\ \ref{fig:gf} are repellors
of the one-loop RG equations.

Let us consider the case with scaling dimensions
$\Delta_1,\Delta_2$ such that $\Delta_1+\Delta_2=d$. It will be convenient
to set $2a \equiv d-2\Delta_1$. The RG equations, whose stability
properties we want to study are
\begin{subequations}
\beq
\label{appCaa}
\dot g_{11}=-8g_{11}^2-8g_{12}^2+2a g_{11}~,
\eeq
\vspace{-.7cm}
\beq
\label{appCab}
\dot g_{22}=-8g_{22}^2-8g_{12}^2-2a g_{22}~,
\eeq
\vspace{-.7cm}
\beq
\label{appCac}
\dot g_{12}=-8g_{12}(g_{11}+g_{22})
~.
\eeq
\end{subequations}
We will perturb around the fixed point
\beq
\label{appCad}
\bar g_{11}=-\bar g_{22}=f~, ~ ~ \bar g_{12}=g~, ~ ~ {\rm with}~~
4f^2-af+4g^2=0
~.
\eeq

Setting
\beq
\label{appCae}
g_{ij}=\bar g_{ij} + h_{ij}
\eeq
we find to leading order in the perturbation $h_{ij}$
\begin{subequations}
\beq
\label{appCaf}
\dot h_{11}=-2(8f-a)h_{11}-16gh_{12}
~,
\eeq
\vspace{-.7cm}
\beq
\label{appCag}
\dot h_{22}=2(8f-a)h_{22}-16g h_{12}
~,
\eeq
\vspace{-.7cm}
\beq
\label{appCai}
\dot h_{12}=-8g(h_{11}+h_{22})
~.
\eeq
\end{subequations}
These equations are more transparent in terms of the
functions
\beq
\label{appCaj}
h_\pm=h_{11}\pm h_{22}~, ~ ~ h=h_{12}
~.
\eeq
A trivial rewriting of eqs.\ \eqref{appCaf} -- \eqref{appCai} gives
\begin{subequations}
\beq
\label{appCak}
\dot h_+=-2(8f-a)h_{-}-32gh
~,
\eeq
\vspace{-.7cm}
\beq
\label{appCal}
\dot h_{-}=-2(8f-a)h_{+}
~,
\eeq
\vspace{-.7cm}
\beq
\label{appCam}
\dot h=-8gh_+
~.
\eeq
\end{subequations}

The solution is
\begin{subequations}
\beq
\label{appCan}
h_+=c_1 e^{2at}+c_2 e^{-2at}~,
\eeq
\vspace{-.7cm}
\beq
\label{appCao}
h_-=c_3- \frac{8f-a}{a} \left (c_1 e^{2at}-c_2 e^{-2at}\right )
\eeq
\vspace{-.5cm}
\beq
\label{appCap}
h=c_4-\frac{4g}{a} \left(c_1 e^{2at}-c_2 e^{-2at} \right )
~.
\eeq
\end{subequations}
where $c_1,c_2,c_3, c_4$ are integration constants. As $t\to +\infty$ (the IR)
there are always directions where the solution is exponentially diverging and
the RG flow drives the theory away from the fixed points on the circle.

\section{The graviton mass matrix for general multi-trace perturbations}
\label{app:genmass}
\def\m{\mu}
\def\n{\nu}
\def\r{\rho}
\def\s{\sigma}

\vspace{-.2cm}
\subsection{A monomial multi-trace deformation}

Consider perturbing $n$ large-$N$ CFTs as follows
\be
S=\sum_{i=1}^n S_i+h\int d^dx ~\Phi\sp \Phi\equiv
\prod_{i=1}^n \OO_i\sp D|\OO_i\rangle=\Delta_i|\OO_i\rangle\sp \sum_{i=1}^n \Delta_i=d
\label{m1}\ee
where $D$ is the dilatation operator. We normalize the scalar single-trace operators
$\OO_i$ so that
\be
\langle \OO_i(x)\OO_i(y)\rangle={1\over |x-y|^{2\Delta_i}}\sp \langle \OO_i|\OO_i\rangle=1
~.
\label{m2}\ee
Defining the stress tensors $T^i_{\mu\nu}$ in the usual manner by the variation
\be
\delta S_i\equiv{1\over 2}\int \sqrt{g}~d^d x ~T^i_{\m\n}~\delta g^{\m\n}
\label{m3}\ee
we find
\be
\d^{\m}T^i_{\m\n}=h~\Phi^i_{\n}\sp \Phi^i_{\n}\equiv \OO_1\cdots \d_{\n}\OO_i\cdots \OO_{n}
\label{m4}\ee
as well as
\be
\langle T^i_{\m\n}|T^j_{\r\s}\rangle=c_i\delta_{ij}\left[\delta_{\m\r}\delta_{\n\s}
+\delta_{\n\r}\delta_{\m\s}-{2\over d}\delta_{\m\n}\delta_{\r\s}\right]
~.
\label{m5}\ee
These relations guarantee that the total stress tensor is conserved
\be
T_{\m\n}=\sum_{i=1}^nT^i_{\m\n}-h\delta_{\m\n} \Phi\sp \d^{\m}T_{\m\n}=0
~.
\label{m6}\ee

We now construct an additional set of $n-1$ spin-2 operators that are orthogonal
to leading order to the total stress tensor $T_{\m\n}$, and orthogonal to the perturbing
operator
\be
\hat T^i_{\m\n}\equiv \sum_{j=1}^n~a_{ij}T^j_{\m\n}+hb_{i}\Phi\sp i=1,2,\cdots,n-1
~.
\label{m7}\ee
Using the correlation
\be
\langle T^i_{\m\n}|\Phi\rangle=h{\Delta_i\over d}\delta_{\m\n}
\label{m8}\ee
we obtain the following conditions
\begin{subequations}
\be
\langle T_{\m\n}|\hat T^i_{\r\s}\rangle={\cal O}(h^2)~~~\Rightarrow~~~
\sum_{j=1}^n~a_{ij}~c_j=0
~,
\label{m9}\ee
\be
\langle \hat T^i_{\m\n}|\hat T^j_{\r\s}\rangle=\hat c_i\delta_{ij}+{\cal O}(h^2)~~~
\Rightarrow~~~\sum_{k=1}^n~a_{ik}a_{jk}~c_k=0
\sp \hat c_i=\sum_{k=1}^n~a_{ik}^2~c_k
~,
\label{m10}\ee
\be
\langle \hat T^i_{\m\n}|\Phi\rangle=0 ~~~\Rightarrow~~~
\sum_{j=1}^n~a_{ij}\Delta_j+d~b_i=0~~~\Rightarrow~~~
b_i=-{1\over d}\sum_{j=1}^n~a_{ij}\Delta_j
~.
\label{m11}\ee
\end{subequations}
Note that since $\sum_{i=1}^n \Delta_i=d$, $\langle T_{\m\n}|\Phi\rangle=0$
automatically. In total, we have $n-1+{1\over 2}(n-1)(n-2)+n-1={1\over 2}(n-1)(n+2)$
equations for the $(n+1)(n-1)$ unknowns $a_{ij}$, $b_i$. An extra $(n-1)$ variables
correspond to the normalizations of the norms $\hat c_i$, and the remaining
${1\over 2}(n-1)(n-2)$ undetermined coefficients correspond to $O(n-1)$ rotations
of the $\hat T^i$ that leave the previous conditions invariant.

We may now use (\ref{m4}) and (\ref{m7}) to obtain
\be
\d^{\m}\hat T^i_{\m\n}= h\sum_{j=1}^n~(a_{ij}+b_i)\Phi^j_{\n}
\label{m12}\ee
Since
\be
\langle \d_{\m}\OO|\d_{\m}\OO\rangle=2\Delta \langle \OO|\OO\rangle~~~
\Rightarrow~~~\langle \Phi^i_{\m}|\Phi^j_{\m}\rangle=2\Delta_{i}\delta_{ij}
\label{m13}\ee
we find the inner products
\be
\langle \d^{\m}\hat T^i_{\m\n}|\d^{\m}\hat T^j_{\m\n}\rangle
=2h^2\sum_{k=1}^n\Delta_{k}(a_{ik}+b_i)(a_{jk}+b_j)
~.
\label{m14}\ee
At the same time, the conformal algebra gives
\be
\langle \d^{\m}\hat T^i_{\m\n}|\d^{\m}\hat T^j_{\m\n}\rangle
={2(d+2)(d-1)\over d}\sqrt{\hat c_i}(\Delta_{ij}-d\delta_{ij})
\sqrt{\hat c_j}
\sp D |\hat T^i_{\m\n}\rangle\equiv \Delta_{ij}|\hat T^j_{\m\n}\rangle
~.
\label{m15}\ee
From this relation and
\be
M^2_{ij}=d(\Delta_{ij}-d\delta_{ij})
\label{m16}\ee
we obtain for the graviton mass matrix
\be
M^2_{ij}={d^2h^2\over \sqrt{\hat c_i\hat c_j}(d+2)(d-1)}
\sum_{k=1}^n\Delta_{k}(a_{ik}+b_i)(a_{jk}+b_j)
~.
\label{m17}\ee

For $n=3$ and the choice $a_{11}=a_{22}=1$, $a_{21}=0$ we obtain
\be
M^2={d^2h^2\over (d+2)(d-1)}\left(\begin{matrix}{(c_1+c_2+c_3)
\Delta_1(d-\Delta_1)\over dc_1(c_2+c_3)}
&-{\Delta_1\over d}{c_3\Delta_2-c_2\Delta_3\over c_2+c_3}
\sqrt{{c_1+c_2+c_3\over c_1c_2c_3}} \\
-{\Delta_1\over d}{c_3\Delta_2-c_2\Delta_3\over c_2+c_3}
\sqrt{{c_1+c_2+c_3\over c_1c_2c_3}}  & {dc_2^2\Delta_1+d(c_2+c_3)^2\Delta_2-
(c_3\Delta_2+c_2(\Delta_1+\Delta_2))^2\over dc_2c_3(c_2+c_3)}\end{matrix}\right)
\label{m18}\ee
with eigenvalues
\begin{subequations}
\be
M^2_{\pm}={d^2h^2\over (d+2)(d-1)}{B\pm \sqrt{B^2-4AC}\over 2A}
~,
\label{m19}\ee
\be
A=dc_1c_2c_3\sp C=(c_1+c_2+c_3)\Delta_1\Delta_2\Delta_3
~,
\label{m20}\ee
\be
B=c_1c_2(\Delta_1+\Delta_2)\Delta_3+c_1c_3(\Delta_1+\Delta_3)\Delta_2
+c_2c_3(\Delta_2+\Delta_3)\Delta_1
~.
\label{m21}\ee
\end{subequations}
When $\Delta_3=0$ we obtain $M_-=0$
and
\be
M^2_{+}=\left({1\over c_2}+{1\over c_3}\right){d\Delta_1\Delta_2~ h^2\over (d+2)(d-1)}
\label{m22}\ee
in agreement with previous calculations.

For $c_1=c_2=c_3=c$, $\Delta_1=\Delta_2=\Delta_3=d/3$ we obtain
\be
M^2_{\pm}={h^2d^3\over 3(d+2)(d-1)~c}
~.
\label{m23}\ee

\subsection{The general case}

The general case involves $n$ CFTs that we could think as nodes in a graph.
For each subset $I$ of $m_I\leq n$ CFTs that possess single trace operators $O^I_i$
with dimensions $\Delta^I_i$ summing up to $d$, we can perturb with an operator
that is the product $\Phi^I=\prod_{i=1}^{m_I} O^I_i$. The upper limits on $m$
depend crucially on the spacetime dimension $d$ through the unitarity bound
for scalar operators. This maximum on $m$ is 2 for dimensions $d\geq 6$, 3 for
dimensions $d=5,4$, 5 for $d=3$ and $\infty$ for $d=2$. Multi-trace operators
with fields entirely in a single theory can also be added to the action but they will
not affect the calculation of the graviton mass since they have zero overlap both
with the stress tensors as well as the perturbing operators.
Hence, we will consider the action
\be
S=\sum_{i=1}^n S_i+\sum_Ih_I\int d^dx \Phi^I
~.
\label{m24}\ee

The total stress tensor is conserved
\be
T_{\m\n}=\sum_{i=1}^nT^i_{\m\n}-\sum_Ih_I\delta_{\m\n} \Phi^I\sp \d^{\m}T_{\m\n}=0
~.
\label{m25}\ee
The remaining $n-1$ spin-2 operators are now
\be
\hat T^i_{\m\n}\equiv \sum_{j=1}^n~a_{ij}T^j_{\m\n}+\sum_I h_I b_{Ii}~\Phi^I\sp i=1,2,\cdots,n-1
~.
\label{m26}\ee
The orthogonality conditions give
\begin{subequations}
\be
\langle T_{\m\n}|\hat T^i_{\r\s}\rangle={\cal O}(h^2)~~~\Rightarrow~~~
\sum_{j=1}^n~a_{ij}~c_j=0
~,
\label{m27}\ee
\be
\langle \hat T^i_{\m\n}|\hat T^j_{\r\s}\rangle=\hat c_i\delta_{ij}+{\cal O}(h^2)~~~
\Rightarrow~~~\sum_{k=1}^n~a_{ik}a_{jk}~c_k=0
\sp \hat c_i=\sum_{k=1}^n~a_{ik}^2~c_k
~,
\label{m28}\ee
\be
\langle \hat T^i_{\m\n}|O^I\rangle=0 ~~~\Rightarrow~~~
\sum_{j=1}^n~a_{ij}\Delta^I_j+d~b_{Ii}=0~~~\Rightarrow~~~
b_{Ii}=-{1\over d}\sum_{j=1}^n~a_{ij}\Delta^I_j
~.
\label{m29}\ee
\end{subequations}

Consequently,
\be
\d^{\m}\hat T^i_{\m\n}=\sum_I~ h_I~\sum_{j=1}^n~(a_{ij}+b_{Ii})\Phi^I_{j,\n}\sp
\Phi^I_{j,\n}\equiv O^I_1\cdots \d_{\n}O^I_j\cdots O^I_{m_I}
~,
\label{m30}\ee
\be
\langle \d^{\m}\hat T^i_{\m\n}|\d^{\m}\hat T^j_{\m\n}\rangle
=2\sum_I h_I^2\sum_{k=1}^n\Delta^I_{k}(a_{ik}+b_{Ii})(a_{jk}+b_{Ij})
\label{m31}\ee
from which we finally obtain the graviton mass matrix
\be
M^2_{ij}={d^2\over \sqrt{\hat c_i\hat c_j}(d+2)(d-1)}\sum_I h_I^2
\sum_{k=1}^n\Delta^I_{k}(a_{ik}+b_{Ii})(a_{jk}+b_{Ij})
~.
\label{m32}\ee

\section{A general class of fixed points in multi-trace deformations}
\def\nh{\hat N}

This appendix was added in January 2009 in order to present explicitly the general case
that we had derived but did not report on in the original version of the paper as it was not directly relevant to our result. We recognize, however, that such results are useful in several other 
applications and we are therefore presenting them here.

We consider $\nh$ not necessarily identical $d$-dimensional CFTs and from each one we 
pick a scalar single-trace operator. The action of the perturbed theory we want to consider is
defined to be
\be
S=\sum_{i=1}^{\nh}S_i+\sum_{i,j=1}^{\nh}g_{ij}~O_iO_j
\label{ap1}\ee
where $S_i$ are the actions of the individual CFTs, $g_{ij}$ are real symmetric
coupling constants and the operator $O_i\in$CFT$_i$ has scaling dimension $\Delta_i$.
Note that this is not the most general multitrace coupling between $\nh$ CFTs.
One can also introduce higher products of operators coupling together more 
than two theories. Such examples have been discussed earlier in this paper.
Even if we remain with perturbations that are quadratic in single-trace operators, 
we have the option of using different operatorsfrom a single theory to participate 
to its couplings with the other theories. For example, the operator coupling CFT$_i$ to 
CFT$_{j\not=i}$ is $O_iO_j$ but the operator coupling CFT$_i$ to CFT$_{k\not=j}$ can
be $\Omega_iO_k$, with $\Omega_i\not= O_i$. We do not consider such couplings but 
we will comment on them at the end.

Now we would like to find all the possible fixed points in the space of theories (\ref{ap1}) 
parametrized by $g_{ij}$. The fixed-point equations generalizing (\ref{bdrgca}-\ref{bdrgcc}) 
are
\be
(d-\Delta_i-\Delta_j)g_{ij}-8\sum_{k=1}^{\nh} g_{ik}g_{jk}=0\sp \forall ~i,j=1,2,\cdots,\nh
\label{ap2}\ee

$(i)$ For generic values of $\Delta_i$ only decoupled fixed points exist. There are 
$2^{\nh}$ of them given by
\be
g_{ij}=0\sp \forall~ i\not=j\sp g_{ii}=0~~~{\rm  or}~~~g_{ii}={(d-2\Delta_i)\over 8}
~.
\label{ap3}  
\ee

$(ii)$ If all $\Delta_i$ are equal, $\Delta_i=\Delta$ $\forall~i$, then a fixed point with 
non-trivial coupling exists. This can be found by diagonalizing $g_{ij}$, using an orthogonal transformation, and then solving the equations. The solutions can be written in the form
\be
g_{ij}=\sum_{k=1}^{\nh}O_{ik}O_{jk}D_k\sp 
\sum_{k=1}^{\nh}O_{ik}O_{jk}=\delta_{ij}\sp D_{k}=0 ~~~{\rm  or}~~~D_k={(d-2\Delta)\over 8}
~.
\label{ap4} 
\ee
Therefore the $\nh\times\nh$ matrix $O$ is orthogonal, and there are $\nh+1$ possible 
diagonal matrices $D$, parametrized by their number of zeros in the diagonal. In particular,
the matrices $D={\bf 0}$ and $D={(d-2\Delta)\over 8}{\bf 1}$ correspond to fixed points that have 
no continuous parameters as the orthogonal matrix of (\ref{ap4}) drops out. The first one is the 
trivial solution while the second is one of the solutions already found in (\ref{ap3}).
The rest of the $\nh-1$ solutions are non-trivial. For example, for $\nh=2$, the nontrivial 
solution can be written in terms of an angle $\theta$ as
\be
g_{11}={(d-2\Delta)\over 8}\cos^2\theta\sp g_{22}
={(d-2\Delta)\over 8}\sin^2\theta\sp g_{12}={(d-2\Delta)\over 16}\sin(2\theta)
\label{ap5}
\ee
and was found earlier in the paper. The general solution where the matrix $D$ has 
$m<\nh$ zeros has continuous parameters isomorphic to the coset $O(\nh)/O(\nh-m)$.

A special case of such fixed points is the replica fixed point found independently in 
\cite{taka}
\be
g_{ij}={d-2\Delta\over 8\nh}M_{ij}\sp M_{ij}=1~~\forall~~ i,j=1,2,\cdots,\nh
~.
\ee

Note that there is no solution of $\Delta={d\over 2}$. In this case the perturbations are 
asymptotically free.

$(iii)$ The system of equations with all $\Delta_i$ equal to $\Delta$ have the following symmetry
\be
\Delta\to d-\Delta\sp g_{ij}\to -g_{ij}
\ee
We may exploit this symmetry to find solutions if a subset of dimensions are $\Delta$ and the 
rest are equal to $d-\Delta$.

$(iv)$ We consider now the case where exactly half of the $\Delta_i$ are equal 
($\Delta_i=\Delta$, $i=1,2,\cdots m$), and the other half are equal to
$d-\Delta$: ($\Delta_i=d-\Delta$, $i=m+1,m+2,\cdots \nh$). For this we  
assume that $\nh=2m$. Then the fixed point equations (\ref{ap2}) become
\be
(d-2\Delta)g_{ij}-8\sum_{k=1}^{\nh} g_{ik}g_{jk}=0\sp \forall ~i,j=1,2,\cdots,m
~,
\label{ap6}
\ee
\be
(d-2\Delta)g_{ij}+8\sum_{k=1}^{\nh} g_{ik}g_{jk}=0\sp \forall ~i,j=m+1,m+2,\cdots,\nh
~,
\label{ap7}
\ee
\be
\sum_{k=1}^{\nh} g_{ik}g_{jk}=0\sp \forall ~i=1,2,\cdots,m~~~~j=m+1,m+2,\cdots,\nh
~.
\label{ap8}
\ee
This system can be solved by diagonalizing first the two matrix blocks, 
$g_{ij}, ~~i,j=1,2,\cdots,m$ and $g_{ij}, ~~i,j=m+1,m+2,\cdots,\nh$.
The fixed-point solution for $g_{ij}$ can be split into 4 $m\times m$ matrices
\be
g=\left(\begin{matrix}
x_1{\bf 1} & x_2 ~O\\ x_2~O^{T} & -x_1{\bf 1}
\end{matrix}\right)\sp x_1={d-2\Delta\over 16}(1+\sin\theta)\sp x_2
= {d-2\Delta\over 16}\cos\theta
\ee
where $O$ is an arbitrary $m\times m$ orthogonal matrix.
The solution has continuous parameters isomorphic to $O(m)\times O(2)$.
The case $m=1$ was treated in the main body of this paper and corresponds to the 
circle of fixed points found there.

$(v)$ Finally we consider the case where there is an unequal number of $\Delta$ and 
$d-\Delta$ eigenvalues. Let $m>\nh-m$ be the number of $\Delta$ eigenvalues.
Then, the non-trivial solution is obtained as follows: up to an overall $O(m)$ transformation 
the solution is composed of the solution in $(iv)$ for the subset of $\nh-m$ $\Delta$ and 
$d-\Delta$, and of solution $(ii)$ for the rest $\nh-2m$ entries.

We are now in a position to describe the algorithm for finding the general fixed point solution.
First we separate the $\nh$ dimensions $\Delta_i$ into groups. Each group contains 
dimensions that are either the same or conjugate. Self-conjugate dimensions 
$\Delta={d\over 2}$ can only lead to trivial fixed points, so they are discarded. We label 
these groups by an index $I$. Fixed point values of the coupling constants cannot mix 
different groups, hence it is enough to focus on a single generic one. The $I$-th group 
is characterized by two natural numbers, $n_+$, $n_-$, that denote respectively the 
number of times that a dimension $\Delta_I$ occurs, and the number of times its 
conjugate $d-\Delta_I$ occurs. The fixed point solution in group $I$ is the one given 
above in $(v)$.

\end{appendix}


\end{document}